\documentclass[aps,roupedaddress,draft,showpacs,showkeys]{revtex4-1}

\usepackage{amssymb}
\usepackage{amsmath}

\begin{document}

%\preprint{}

\title{A Process Algebra Approach to Quantum Mechanics}


\author{William H. Sulis}
\affiliation{McMaster University and The University of Waterloo}
\email{sulisw@mcmaster.ca}

\date{\today}

\begin{abstract}
The process approach to NRQM offers a fourth framework for the quantization of physical systems. Unlike the standard approaches (Schr\"odinger-Heisenberg, Feynman, Wigner-Gronewald-Moyal), the process approach is \emph{not} equivalent to NRQM and is \emph{not} merely a re-interpretation. The process approach provides a dynamical \emph{completion} of NRQM. Standard NRQM arises as a asymptotic quotient by means of a set-valued process covering map, which links the process algebra to the usual space of wave functions and operators on Hilbert space. The process approach offers an emergentist, discrete, finite, quasi-non-local and quasi-non-contextual realist interpretation which appears to resolve many of the paradoxes and is free of divergences. Nevertheless, it retains the computational power of NRQM and possesses an emergent probability structure which agrees with NRQM in the asymptotic quotient. The paper describes the process algebra, the process covering map for single systems and the configuration process covering map for multiple systems. It demonstrates the link to NRQM through a toy model. Applications of the process algebra to various quantum mechanical situations - superpositions, two-slit experiments, entanglement, Schr\"odinger's cat - are presented along with an approach to the paradoxes and the issue of classicality.
\end{abstract}

\pacs{03.65.Ta, 03.65.Ud, 02.10.De, 02.30.Px, 02.40.Ul, 02.50.Le}
\keywords{process theory, quantum foundations, discrete models}

\maketitle
\section{Introduction}

There have been three highly successful approaches to the quantization of physical systems. Each may be thought of as a generalization of a particular perspective on classical mechanics. The most highly discussed   is that of Schr\"odinger and Heisenberg, which is based upon the Hamiltonian formulation of classical mechanics \cite{vonNeumann}. It utilizes the mathematics of functional analysis on a Hilbert space whose functions are defined on an abstract configuration  space. The second most widely described approach is that of Feynman, which is based on the Lagrangian formulation of classical mechanics \cite{Feynman}. It utilizes the mathematics of path integrals defined on spacetime (and more often on configuration space). The third approach is less well known and can be attributed to Weyl, Wigner, Gronewald and Moyal, and is based on the Poisson bracket formulation of classical mechanics \cite{Curtright}. It uses the mathematics of $\bigstar$ products and Wigner functions defined upon a classical phase space. Although differing in their formulation, these three approaches have been shown to be equivalent to one another.

In spite of their success from mathematical and computational perspectives, it is fair to say that none of these approaches have provided quantum mechanics with an internally consistent ontological interpretation free of paradoxes. It is often argued that such an interpretation is unnecessary but the persistent tension between quantum mechanics and special relativity, especially in the setting of entanglement, the inability to achieve a consistent formulation of quantum gravity, the subtle discrepancies between observation and the standard model, the inability (to date) to find experimental support for extensions of the standard model, the near tour de force of effort required to eliminate divergences in quantum field theory, all suggest that in spite of its success it remains possible that our current models of quantum mechanics are not exactly correct. It is quite possible that something subtle has been missed in either our mathematical or our conceptual formulations of quantum mechanics. 

There has been much effort expended over the years to find alternative mathematical frameworks for quantizing physical systems but few if any have garnered much support: Bohmian mechanics \cite{Bohm2}, Wolfram's cellular automata \cite{Wolfram}, continuous spontaneous localization (CSL) \cite{Ghirardi}, Finkelstein's quantum relativity \cite{Finkelstein1,Finkelstein}, Noyes's bit-string physics \cite{Noyes}, Bastin and Kilmister's combinatorial physics \cite{Bastin}, Hiley's process physics \cite{Hiley}, Cahill's process physics \cite{Cahill}, Bodiyono's fluctuation model \cite{Bodiyono, Bodiyono1, Bodiyono2, Bodiyono3}.  There have been attempts to formulate quantum mechanics using different conceptual frameworks such as quantum information \cite{Chiribella}, quantum logic \cite{Beltrametti} and category theory \cite{Coecke}. Quantum information is the most actively pursued at the present time but much more work is needed in order to be able to \emph{derive} quantum mechanics from quantum information. Bohmian mechanics has also been extensively studied and provides the most realist interpretation but requires an extreme form of non-locality and struggles in the context of quantum field theory.

This paper presents a new approach to (non-relativistic) quantum mechanics based upon the concept of process as developed originally by Whitehead \cite{Whitehead,Shimony,Eastman} and using the mathematics of process algebra \cite{Sulisthesis}. The process algebra model offers a fourth approach to the quantization of physical systems which provides an entirely novel mathematical framework, a consistent ontological interpretation based in spacetime, and which appears to be free of the usual paradoxes and  divergences. Moreover, the process algebra approach is \emph{not} equivalent to quantum mechanics. It is a larger mathematical framework which represents dynamical information lacking in the above formulations of quantum mechanics. Quantum mechanics arises as a quotient operation in the context of a suitable continuum approximation and thus the process model provides a form of dynamical completion of quantum mechanics while still giving rise to the usual probabilistic structure of quantum mechanics.

The process approach is based upon research in complex systems theory \cite{Sulisad, Sulisct, Sulisjmp} and non-Kolmorogov probability theory \cite{Gleason,Khrennikov,Khrennikov1,vonNeumann}. The latter has shown that non-Kolmogorov probability structure appears in the classical world as well as in quantum mechanics. This realization has offered new insights \cite{Sulisthesis}  into the on-going debates about whether  reality is local or non-local \cite{Maudlin1,Bohm1,Aharonovp, Greenstein,Larsson}, contextual or non-contextual \cite{Kochen, Leggett, Mermin, Pusey}  and whether the wave function is real or merely heuristic \cite{Colbeck2,Ney, Spekkens,Barrett1,Colbeck1,Ord}.

The key idea of the process approach is to turn the usual reductionist argument on its head and to ask to what extent top-level concepts of complex theory and emergence apply at the lowest levels  \cite{Sulist, Sulistn,Sulisci,Trofimova}. The notion of a generated spacetime was inspired by  Sorkin's causal set programme \cite{Sorkin} but diverges from it in its implementation. The necessity to consider discrete models appears to be forced upon us by recent work by Gisin \cite{Gisin}, who constructed a Bell type inequality showing that either one must reject the principle of continuity or accept instantaneous information transfer between space-like separated entities (thus rejecting special relativity). The key idea for recovering (at least the illusion of) continuity is to be found in the work of Kempf \cite{Kempf}. 

The process model  presents a decidedly unromantic (to quote Bell \cite{Bell}) model of quantum mechanics in which wave functions correspond to real physical waves, space-time and physical entities are emergent, and which is discrete, finite, intuitive, causal, quasi-local and quasi-non-contextual, yet retains the computational power of standard quantum mechanics. The model considers a discrete, finite causal space which is \emph{generated} rather than simply existing. The physics plays out entirely on this causal space using only causally local information. Non-relativistic quantum mechanics (NRQM) appears as an idealization when information can be considered to be infinite and the spacing between elements of the causal space to be infinitesimal.

\section{Informons}

In his Process Theory, Whitehead \cite{Whitehead, Shimony,Eastman} posited that the elements of physical reality do not simply \emph{exist}, but rather are \emph{emergent} upon a lower level of entities which he called \textquotedblleft actual occasions\textquotedblright . Here they shall be referred to as \emph{informons} (short for informational monads) to reflect their fundamental informational character. Informons are thought of as passive carriers of information which are generated moment to moment through the actions of processes, which interpret, transform and supplement the information of the current generation of informons and incorporate it into the next.
 Informons are postulated to be discrete, delocalized, finite, and organized into distinct generations. They are transient - arising, persisting briefly as their information is incorporated into the next generation of informons, then abating, much like the tokens used to play a game or the pixels on a computer screen. Physical entities appear at the emergent level analogous to the image patterns that form on such a screen. Information passes causally from one generation to the next, \emph{never} within a generation. As a consequence, special relativity is not violated \cite{Sulisct}.

Informons have no dynamics but they do possess form. Mathematically each informon $n$ is assigned a tuple $\mathbf{p}_{n}$ of properties inherited from the process $\mathbb{P}$ that generates it. The most important of these is the \textquotedblleft strength \textquotedblright\, or \textquotedblleft coupling effectiveness \textquotedblright\, $\Gamma_{n}$ of the generating process $\mathbb{P}$ as determined at the informon $n$. Here, coupling refers to interactions between $\mathbb{P}$ and other processes. The coupling effectiveness expresses the compatibility between informons of different processes and therefore their likelihood of interacting. This notion of compatibility arises in complex system theory and was first proposed by Trofimova \cite{Trofimova} in her work on ensembles with variables structures. The strength $\Gamma(n)$ enters into the construction of the wave function but it does \emph{not} determine the probability of occurrence of informons. Rather, it enters into the determination of  the probability that its generating process will couple with a measurement process and ultimately yield a measurement value. The probability associated with measurements thus becomes another emergent feature of the process viewpoint \cite{Sulisthesis,Sulisjmp,Sulis2}.

Informons arise and abate but their information propagates  in a manner akin to a dissipative wave. Following an idea of Markopoulou \cite{Markopoulou},  prior information is associated with the current informon in the form of a content set $G_{n}$. $G_{n}$  consists of all of those informons from previous generations that pass information to $n$. The passing of information from one generation to another induces a causal structure between generations. Given two informons $m,m'$ in $G_{n}$, write $m\rightarrow m'$ (or $m<m'$) if $m$ was in a earlier generation than $m'$ and information was passed from $m$ to $m'$.  $G_{n}$ is thus an acyclic directed graph whose vertices consist of causally related prior informons. Equivalently it may be viewed as a partially ordered set. Note that $m<n$ for every $m\in G_{n}$. Furthermore, if $\mathcal{I}$ represents a single generation of informons, then $\mathcal{I}$ forms an antichain and $\cup_{n\in \mathcal{I}} G_{n}$ also forms an acyclic directed graph, ensuring consistency of the causal structure. 

Since one generation $\mathcal{I}$ of informons represents an instance of reality, the information upon which the process dynamics depends must therefore lie within $\{\Gamma_{n},\mathbf{p}_{n}|n\in \mathcal{I}\}$ and $\{G_{n}|n\in \mathcal{I}\}$.

In many cases it is also useful to assume that there is an indefinite metric $d$ assigned to $L_{\mathcal{I}}=\mathcal{I} \cup_{n\in \mathcal{I}} G_{n}$ such that for any $m,m'\in K$, $d(m,m')>0$ iff $m<m'$. This assigns a \textquotedblleft time-like\textquotedblright\, distance between successive generations and a \textquotedblleft space-like\textquotedblright\, distance within a generation and between informons of distinct generations that are causally unrelated.

It is useful to exploit the artifice of  history. If $\mathcal{I}_{n}$ is the current generation, let $\mathcal{I}_{<n}=\cup_{k}\mathcal{I}_{k}$, $k<n$ (note we allow $k<0$) and $G_{<n}=\cup_{m}G_{m}, n\in \mathcal{I}_{n}$. $\mathcal{I}_{<n}$ represents complete information about the past and is a mathematical convenience, not an ontological model. $G_{<n}$ on the other hand represents past information that is accessible in the current generation.

Following the ideas of archetypal dynamics \cite{Sulisad}, the information embodied in informons must be interpreted by both the processes acting upon it and by observers attempting to measure and understand it. The interpretation should connect informons to some physical theory and its models. Here the interpretation will utilize  causal manifolds (a causal manifold is a generalization of the light cone structure of Minkowski space to arbitrary manifolds \cite{Borchers}) and Hilbert spaces over such manifolds). 
 
Each informon $n\in \mathcal{I}_{k}$ is therefore interpreted as a point $\mathbf{x}_{n}$ in a causal manifold $\mathcal{M}$ with causal ordering $\prec$ and metric $\rho$. Given $n,n'$, if $n<n'$ then $\mathbf{x}_{n}\prec \mathbf{x}_{n'}$. Each generation $\mathcal{I}_{n}$ is thus interpreted in the causal manifold $\mathcal{M}$ as a discrete sampling $\{\mathbf{m}_{k}, k\in \mathcal{I}_{n}\}$ of a spacelike hypersurface $\mathcal{M}_{\mathcal{I}_{n}}$. The informons of $\mathcal{I}_{n}$ constitute a causal antichain in both $\mathcal{I}_{n}$ and $\mathcal{M}$.  Sometimes one requires that $\rho (n,n')=d(n,n')$ or that $\rho(n,n')= d(n,n')\pm\epsilon$ for a small error $\epsilon$. The wave function $\Psi(\mathbf{z})\in\mathcal{H}(\mathcal{M})$ of a physical entity (including the vacuum) is understood to represent  a physical wave defined on the causal manifold $\mathcal{M}$ (or sometimes a space-like hypersurface of $\mathcal{M}$). Each informon is  interpreted as providing a local Hilbert space contribution to this wave function of the form $\phi_{n}(\mathbf{z})=\Gamma_{n}f_{n}(\mathbf{z},\mathbf{x}_{n})$, such that $\Psi(\mathbf{z})=\sum_{n\in \mathcal{I}_{k}}\phi_{n}(\mathbf{z})$. Where definable, $f_{n}$ will be a translation of a single generating function $g$, that is $f_{n}(\mathbf{z},\mathbf{x}_{n})=T_{\mathbf{x}_{n}}g(\mathbf{z})=g(\mathbf{z}-\mathbf{x}_{n})$.
For example one might take $g(\mathbf{z})=\sin \omega \mathbf{z}/\omega \mathbf{z}$. The informons thus form a discrete set of generators $\{\phi_{n_{i}}(\mathbf{z})\}$ of a wave function $\Psi(\mathbf{z})$ defined on $\mathcal{M}$ by $\Psi(\mathbf{z})\approx\sum_{n_{i}}\phi_{n_{i}}(\mathbf{z})$. The restriction of $\Psi$ to $\mathcal{M}_{\mathcal{I}_{n}}$ is denoted $\hat\Psi$. The pair $(\mathbf{x}_{n},\phi_{n}(\mathbf{z}))$ is called the \emph{interpretation} of $n$, $\mathbf{x}_{n}$ the local $\mathcal{M}$-interpretation and $\phi_{n}(\mathbf{z})$ the local $\mathcal{H}(\mathcal{M})$-interpretation. Note that $\Gamma_{n},\mathbf{p}_{n},G_{n}$ are \emph{intrinsic} features of an informon while the interpretations $\mathbf{m}_{n},\phi_{n}(\mathbf{z})$ are \emph{extrinsic} features. Note that the generation of informons does \emph{not} depend upon the interpretation. This is referred to as frame independence. 

An informon is denoted as $[n]<\alpha\ _{n}>\{G_{n}\}$ where $\alpha_{n} = (\mathbf{x}_{n},\phi_{n}(\mathbf{z}),\Gamma_{n},\mathbf{p}_{n})$. In order to ensure that the causal structure of a generation remains consistent additional constraints are necessary. The formal mathematical structure of a generation of informons is actually that of a \emph{causal tapestry} \cite{Sulisct}. The details are not necessary for this paper and the interested reader is referred to the literature. The terms generation and causal tapestry will be used interchangeably. 

\section{Process}

Process is considered to be the generator of informons. Processes are \emph{not} situated \emph{in} spacetime since they \emph{generate} spacetime. A process may be \emph{active} in which case it acts in a series of rounds to generate a coherent collection of informons, or it may be \emph{inactive}. Interactions determine whether an active  process becomes inactive or vice-versa. A process $\mathbb{P}$ is described by several parameters: one or more tuples of properties ($\mathbf{p}\in D$) (such as mass, charge, spin, energy, angular momentum, linear momentum, lepton number etc), the number of informons generated during a round ($R$), the number of previous informons whose information is incorporated into a nascent informon ($r$), the number of rounds needed for a generation ($N$), the temporal and spatial scales of informons ($t_{P},l_{P}$) (which may or may not be linked to individual properties). Moroever, a process is alos described by the strategy that it uses to generate informons.

A complete action of a process creates a single generation of informons. 
A system unfolds as a succession of generations.

\begin{displaymath}
\cdots\mathcal{I}_{n}\stackrel{\mathbb{P}_{n}}{\rightarrow}\mathcal{I}_{n+1}\stackrel{\mathbb{P}_{n+1}}{\rightarrow}\mathcal{I}_{n+2}\cdots
\end{displaymath}

\noindent The triple of current generation $\mathcal{I}_{n}$, process $\mathbb{P}_{n}$, and nascent generation $\mathcal{I}_{n+1}$, forms what philosophers term a \emph{compound present}.

A \emph{primitive} process is defined as generating a single informon during a single round ($R=1$). It represents a single physical entity such as a single photon or electron. A multiple entity process generates multiple informons during each round, i.e. $R>1$.  Multiple entity processes are constructed from primitive processes using the process algebra as discussed in the next section. A primitive process generates, one by one, a succession of informons $n_{1},n_{2},\ldots $, which taken in totality form a single generation $\mathcal{I}_{n}$. The sequential generation of informons by a process is a key feature of the model. This discrete sequential generation of informons is essential for ensuring that the measurement process possesses both quantized and continuous aspects and for resolving most of the paradoxes. 

 The coupling factor of the process $\mathbb{P}$ at the informon $n_{i}$ is defined as $\Gamma_{n_{i}}=\phi_{n_{i}}(\mathbf{m}_{n_{i}})$ and the strength there is defined as $l_{P}^{3}\Gamma_{n_{i}}^{*}\Gamma_{n_{i}}$. The relative strength, $||\mathbb{P}||_{\mathcal{I}}$, of $\mathbb{P}$ over the causal space $\mathcal{I}$  is defined as $||\mathbb{P}||_{\mathcal{I}}^{2}=\sum_{n_{i}\in \mathcal{I}}l_{P}^{3}\Gamma_{n_{i}}^{*}\Gamma_{n_{i}}$. 
Processes are posited to act \emph{non-deterministically}, a term borrow from computation theory to mean that actions are described by set-valued maps without any intrinsic probability structure. Probabilities are emergent through interactions with other processes, especially measurement processes.

Properties are associated with individual processes which then impose these upon the informons that they generate. Conservation laws and symmetries describe constraints on these properties and thus apply to processes and especially serve to constrain their interactions with other processes. 

The relationship between the action of process and conceptions of time is an open problem. Process action may be considered as being outside of time, and generating time as we experience it. Process action may simply occur at a time scale too small to ever be observable (the section on errors presents one example of this scenario). Process action may occur in a separate time such as arises in two time physics \cite{Bars} or stochastic quantization \cite{Parisi}. 

\section{A Concrete Example: Free Path Integral Strategy}

Before proceeding to discuss the process algebra in more detail, let us consider a simple model designed to provide an in-principle demonstration of the process approach. We will consider a specific representation of the process algebra as a combinatorial game algebra and a specific game strategy, the free path integral strategy. This will illustrate how NRQM, at least in the case of a scalar particle, appears as an asymptotic limit of infinite informons and infinite information transfer. This demonstrates the assertion that quantum mechanics is to be viewed as an idealization, and thus a special case of the more general process model.

A particularly useful heuristic representation of process is as a two player, co-operative, combinatorial game \cite{Conway}, based on the forcing games used in mathematical logic to generate models \cite{Hodges}. Combinatorial games are distinct from economic games. Combinatorial games derive their power and significance from their rich combinatorial and algebraic structure and involve the application of various \emph{tokens} to different sites. A move alters the tokens associated with a position. Most common are games involving two players who possess a distinct set of tokens and strategies to move and who alternate in making moves. Moves are made non-deterministically, meaning that from any position more than one move is possible and the move is freely chosen without any preassigned probability. Games have a definite end either in the form of some set of \emph{} configurations or by reaching a predetermined limit of individual moves. Usually the last player to move is said to \emph{win} the game.  Games are often described by their game tree. A complete play is a path through the game tree beginning with some initial configuration and ending with a final configuration. The algebra of combinatorial games \cite{Conway1} is isomorphic with the process algebra described in later sections. It is customary in the study of combinatorial games to assign each player a strategy which consists of a set of additional rules determining their moves.  In the process case, different strategies may be thought of as representing different dynamics. The notion of weak-epistemic equivalence (defined below) is an important as a tool for separating out classes of strategies for study. For NRQM one is interested in weak-epistemic equivalent strategies capable of generating global wave functions that satisfy particular Schr\"odinger equations.  The free strategy described below is not meant to be definitive but rather to provide an in-principle demonstration of the validity of the process approach. 

For simplicity, consider a single non-relativistic scalar particle with mass $\mathbf{m}$, in a single eigenstate of the Hamiltonian, and interacting with a potential $V(\mathbf{z})$, which summarizes the effect of the environment. The particle is modeled as a process while the environment process is ignored (being incorporated into $V$). This simplification allows the value of $V$ associated with an informon $n$ to simply be a property of the informon. Since the setting is non-relativistic one can set $\mathcal{M}=\mathbb{R}^{4}$ and take the causal order to be given by $(t_{1},x_{1},y_{1},z_{1})\prec (t_{2},x_{2},y_{2},z_{2})$ iff $t_{1}<t_{2}$.

Player I propagates information forward to the nascent generation while Player II uses this to construct the new informons. Let $\mathcal{I}_{n}$ be the current generation and $\mathcal{I}_{n+1}$ the nascent generation.  The causal tapestry will be constructed as a sublattice of a uniform lattice.  Informons may thus be labelled by their lattice site value, which will be of the form $l_{p}=(mt_{P},il_{P},jl_{P},kl_{P})$ where $m, t_{P},l_{P}$ are fixed, $m$ referring to the generation number, $t_{P},l_{P}$ being the lattice spacings, and $-\infty\leq i,j,k\leq \infty$. An informon will be referred to interchangeably by either its label $n$  or its site $l_{n}$. The causal distance between $n\in \mathcal{I}_{m}, n'\in \mathcal{I}_{m+1}$ is given by the Euclidean metric applied to the lattice site values. Hence if $l_{n}=(mt_{P},il_{P},jl_{P},kl_{P})$ and $l_{n'}=((m+1)t_{P},i'l_{P},j'l_{P},k'l_{P})$ then $d(n,n')^{2}=d(l_{n},l_{n'})= t_{P}^{2}+((i-i')^{2}+(j-j')^{2}+(k-k')^{2})l_{P}^{2}$. Note that one could easily set $\mathcal{M}=\mathbb{M}^{4}$, the 4-dimensional Minkowski space with causal order given via the Minkowski metric to obtain the relativistic case.

The process strength $\Gamma_{n}$ of an informon $n$ is thought of as propagating forward to subsequent informons as a discrete (possibly dissipative) wave. The contribution to the next informon $n'$ will depend upon the causal distance between them. If the Lagrangian for the particle is $\mathcal{L}=\frac{p^{2}}{2\mathbf{m}}+V$ and $\mathcal{L}(n,n')=\frac{\mathbf{m}d(n,n')^{2}}{2t_{P}^{2}}+V(n)$. Then each contribution will take the form $e^{(i/\hbar)\mathcal{L}(n,n') t_{P}}$ or alternatively, $e^{(i/\hbar)\mathcal{L}(l_{n},l_{n'})t_{P}}$. Let . 
Note that $\mathcal{L}(n,n')$ does not depend upon the causal manifold or $\mathcal{H}(\mathcal{M})$-interpretations. The physics  takes place solely on the causal tapestry.

The $\mathcal{H}(\mathcal{M})$-interpretation is constructed by means of sinc interpolation \cite{Zayed,Jorgensen,Marks}. Sinc interpolation requires the use of a lattice embedding into $\mathcal{M}$, which admittedly is unrealistic, but Maymon and Oppenheim \cite{Maymon} have shown that non-uniform embeddings will still provide a highly accurate approximation. A more realistic model would require the use of non-uniform embeddings and more sophisticated interpolation techniques, such as Fechtinger-Gr\"ochenik theory \cite{Zayed}. The  physics is independent of the interpolation scheme which merely serves as a bridge to NRQM.

Sinc interpolation depends upon the values of $t_{P}$ and $l_{P}$ and is limited to band-limited functions, meaning that their Fourier transforms have support within the bounded region $[-\sigma,\sigma]$. This class is smaller than $L^{2}(\mathcal{H}(M))$ but possesses a natural ultraviolet cutoff and $L^{2}(\mathcal{H}(M))$ can be approximated in the limit $t_{P},l_{P}\rightarrow 0$.

A basic strategy which provides an in-principle demonstration of the power of the method is the Bounded Radiative Uniform Sinc Path Integral Strategy $(\mathfrak{P}\mathfrak{I}$). The path integral strategy is specified by the parameters $R,r,N,t_{P},l_{P}$ and by 

\begin{enumerate}
\item $\Delta$ (distance bound): arbitrary, determines maximum causal distance of information transmission.
\item $\rho$ (approximation measure): arbitrary, set by the observer according to mathematical or experimental considerations.
\item  $\delta$ (approximation accuracy): arbitrary but bounded by experimental measurements
\item $\omega$ (band limit frequency): bounded by upper limits of energy and momentum of the quantum system
\item $\mathcal{L}$ (Lagrangian): determined by the particulars of the quantum system
\item $p$ (set of properties): here energy, momentum
\end{enumerate}

The informons of the causal tapestry $\mathcal{I}_{n}$ will be embedded into a sub-lattice of the space-like hyper-surface (or time slice) $\{nt_{P}\}\times \mathbb{R}^{3}$ in $\mathcal{M}$. The embedding lattice in $\mathcal{M}$ will thus take the general form $(nt_{P},il_{P},jl_{P},kl_{P})$ for integers $n,i,j,k$. The embedding point in $\mathcal{M}$ of an informon $n$ will be denoted $\mathbf{m}_{n}$. For simplicity, set $l_{n}=\mathbf{m}_{n}$.

The path integral strategy for a single short round proceeds as follows:

\begin{enumerate}
\item Player I moves first. Player I non-deterministically chooses any informon $[n]<\alpha_{n}>\{G_{n}\}=[l_{n}]<(\mathbf{m}_{n},\phi_{n},\Gamma_{n},V(n))>\{G_{n}\}$ from the current tapestry $\mathcal{I}_{m}$ which has not previously been played in this round, where $\mathbf{m}_{n}$ is the $\mathcal{M}$ embedding, $\phi_{n}(\mathbf{z})$ is the local Hilbert space contribution, $l_{n}$ is a lattice site, $\Gamma_{n}$ is the strength of the generating process at $n$ and $V(n)$ is the value of the potential at $n$ .

\item If there is an informon $[n']<\alpha_{n'}>\{G_{n'}\}$ currently in play in the new tapestry $\mathcal{I}_{m+1}$ then Player II tests whether $d(n,n')<\Delta$ in the new tapestry $\mathcal{I}'$. If the bound is exceeded, play reverts back to step I, otherwise it proceeds. If there is no current informon then Player II chooses a label $n'$ not previously used and selects a lattice site $((m+1)t_{P},i'l_{P},j'l_{P},k'l_{P})$ not previously used such that $d(l_{n},l_{n'})<\Delta$ and creates a new informon $[n']<\alpha_{n'}>\{\}$. 
\item Player I next updates the content set. If the new inform already possesses a content set $G_{n'}$, then Player I replaces $G_{n'}$ with $G_{n'} \cup \hat G_{n} \cup \{[n]<\alpha_{n}>\{G_{n}\}\}$ ($\hat G_{n}$ is an order theoretic up-set of $G_{n}$) and checks to ensure that all necessary order conditions are satisfied. If the new informon is nascent, then Player I simply sets $G_{n'} = \hat G_{n} \cup \{[n]<\alpha_{n}>\{G_{n}\}\}$. The content set determines what prior information is permitted in constructing tokens. It only includes informons from the past causal cone of the informon. In the case of NRQM, it turns out that only informons from the current tapestry are needed since the relevant information is already incorporated into their $\mathcal{H}(M)$-interpretations. Thus it suffices if $G_{n'}$ or $\emptyset$ is replaced with $G_{n'} \cup \{[n]<\alpha_{n}>\{G_{n}\}\}$ or $\{[n]<\alpha_{n}>\{G_{n}\}\}$ respectively. 
\item  Player II next determines the causal manifold embedding. If the nascent informon $n'$ already possesses a causal manifold embedding, then Player II does nothing. Otherwise Player II sets $\mathbf{m}_{n'}=\l_{n'}$ and the nascent informon becomes $[n']<\mathbf{m}_{n'},,,V(\mathbf{m}_{n'})>\{G_{n'}\}$.

\item Player I next constructs a token representing the information passing from $n$ to $n'$ and to be used to form the local Hilbert space contribution at $n'$. Denote this token as $\mathcal{T}_{n'n}$. Let $\tilde S[n',n]= \mathcal{L}(n,n')t_{P}$. Let $T_{n}$ denote the set of tokens on $n$. Let $\Gamma_{n}$ denote the sum of the tokens on $n$, that is $\Gamma_{n}=\sum\{\mathcal{T}_{nm}|\mathcal{T}_{nm}\in T_{n}\}$. In what follows $\Phi_{n}(\mathbf{z})$ will refer to the local $\mathcal{H}(\mathcal{M})$-interpretation of $n$.  The relationship between these two is $\Gamma_{n}=(1/A^{3})\Phi_{n}(\mathbf{m}_{n})$, where $A$ is the path integral normalization factor described by Feynman and Hibbs \cite{Feynman} which is appropriate to the current Lagrangian and initial and boundary conditions. The reason for this will become apparent later.  Define the propagator $P_{n'n}=(l_{P}^{3}/A^{3})e^{i\tilde S[n',n]/\hbar}$. Then Player I places a token $\mathcal{T}_{n'n}=P_{n'n}\Gamma_{n}$ on the site $l_{n}$. If there already is a set $T_{n'}$ of tokens on informon $n'$ then replace it by $T_{n'}\cup\{T_{n'n}\}$.

\item Finally Player II must determine the $\mathcal{H}(\mathcal{M})$-interpretation. Set $K(\lambda,x)=\pi x/\lambda$ , $sinc \:x =\sin x/x$. If $\mathbf{z}=(t,x,y,z)$ and $\mathbf{m}_{n'} = ((nt_{P},ml_{P},rl_{P},sl_{P}))$ define 

$$T_{\mathbf{m}_{n'}}sinc_{t_{P},l_{P}}(\mathbf{z})=T_{nt_{P}}sinc(K(t_{P},x))T_{ml_{P}}sinc(K(l_{P},y))\times$$
$$T_{rl_{P}}sinc(K(l_{P},y))T_{sl_{P}}sinc(K(l_{P},z))$$

Player II constructs the $\mathcal{H}(\mathcal{M})$-interpretation by coupling the tokens on the site to a suitable interpolation function, which in the current strategy utilizes a sinc function given as $A^{3}T_{\mathbf{m}_{n}}sinc_{t_{P}l_{P}}(\mathbf{z})$. If the new informon has just been formed, then the $\mathcal{H}(\mathcal{M})$-interpretation is given as $\Phi_{n'}(\mathbf{z}) = \mathcal{T}_{n'n}A^{3}T_{\mathbf{m}_{n'}}sinc_{t_{P},l_{P}}(\mathbf{z})$. If the informon already possesses a $\mathcal{H}(\mathcal{M})$-interpretation, $\Phi_{n'}(\mathbf{z})$,  replace it by the new $\mathcal{H}(\mathcal{M})$-interpretation $\Phi_{n'}(\mathbf{z})+\mathcal{T}_{n'n}A^{3}T_{\mathbf{m}_{n'}}sinc_{t_{P},l_{P}}(\mathbf{z}).$

In other words, add the new token to the collection, sum the token values and couple the sum to the interpolation wavelet. 

\item If no further tokens can be added (either no other contributing sites exist or an external limit has been reached), the round ends and a new round begins. The completed informon is $[n']<\mathbf{m}_{n'},\phi_{n'}(\mathbf{z}),\Gamma_{n'},V(\mathbf{m}_{n'})>\{G_{n'}\}$ where $\Gamma_{n'}=\sum_{n\in \mathcal{I}_{n}}\mathcal{T}_{n'n}$ and $\Phi_{n'}(\mathbf{z})=\Gamma_{n'}A^{3}T_{\mathbf{m}_{n'}}sinc_{t_{P},l_{P}}(\mathbf{z})$.
\end{enumerate}

Play continues until the allotted number of game steps has been reached. At the end of play a new causal tapestry $\mathcal{I}_{m+1}'$ has been created and the old causal tapestry $\mathcal{I}_{m}$ is eliminated, formally becoming a part of $\mathcal{I}^{p}_{m+1}$, the collection of prior tapestries. Any relevant information from $\mathcal{I}_{m}$ now resides within the content sets of the informons of $\mathcal{I}_{m+1}$. Let $n'$ denote an informon of $\mathcal{I}_{m+1}$. Let $L_{n'}$ denote the set of all informons from $\mathcal{I}_{n}$ that form vertices in $G_{n'}$.  The local $\mathcal{H}(\mathcal{M})$-interpretation of $n'$ may now be written as 
$\Phi_{n'}(\mathbf{z})= \sum_{n\in L_{n'}}\mathcal{T}_{n'n}A^{3}T_{\mathbf{m}_{n'}}sinc_{t_{P},l_{P}}(\mathbf{z})=\Gamma_{n'}A^{3}T_{\mathbf{m}_{n'}}sinc_{t_{P},l_{P}}(\mathbf{z}).$

The global $\mathcal{H}(\mathcal{M})$-interpretation on $\mathcal{M}$ is formed by summing the local contributions over all of $\mathcal{I}_{m+1}$, that is $\Phi^{m+1}(\mathbf{z})=\sum_{n'\in \mathcal{I}_{m+1}}\Phi_{n'}(\mathbf{z})$. One may restrict this to the $t=(m+1)t_{P}$ hyper-surface, obtaining, as will be shown below, a highly accurate approximation to the standard quantum mechanical wave function on the hyper-surface. Note that fixing $t=m+1$ causes the time based sinc term to take the value 1 and one indeed obtains a function on the hyper-surface. This approximation will be less accurate when extended to the entirety of $\mathcal{M}$. To achieve greater accuracy requires either summing over the content sets of $\mathcal{I}_{m+1}$, i.e. $\Phi^{m+1,c}_{n'}(\mathbf{z})= \sum_{n\in G_{n'}, n'\in I_{n'}}\Phi_{n}(\mathbf{z})$ or over all of $\mathcal{I}_{m+1}\cup \mathcal{I}_{p}$, $\Phi^{m+1,p}_{n'}(\mathbf{z})= \sum_{n\in \mathcal{I}_{m+1}\cup I_{p}}\Phi_{n}(\mathbf{z})$.

A
\section{Formal Proof of Emergent NRQM}

In the previous section the assertion was made that NRQM can be viewed as an effective theory arising in the asymptotic limit as $N,r\rightarrow \infty$ and $t_{P},l_{P}\rightarrow 0$. To prove this consider the following.

Assume that the particle is generated by a primitive process in an eigenstate of its Hamiltonian. Let $\mathcal{I}_{0}$ denote the initial generation for the particle process $\mathbb{P}$ and assume that on this generation the process strengths $\Gamma_{n}$ correspond to the values of the wave function sampled at the embedding points, i.e $\Gamma_{n}=\Psi(\mathbf{m}_{n})$. 

Parzen's theorem states that if $f(t_{1},\ldots, t_{N})$ is a function band limited to the $N$-dimensional rectangle $B=\prod_{i=1}^{N} (-\sigma_{i},\sigma_{i})$, $\sigma_{i}>0$, $i=1,\ldots,N$ so that its Fourier transform $F(\omega_{1},\ldots,\omega_{N})$ is such that

\begin{displaymath}
\int_{-\sigma_{1}}^{\sigma_{1}}\cdots \int_{-\sigma_{N}}^{\sigma_{N}}|F(\omega_{1},\ldots,\omega_{N})|^{2}d\omega_{1}\cdots d\omega_{N}<\infty,
\end{displaymath}

\noindent $F(\omega_{1},\ldots,\omega_{N})=0$ for $|\omega_{x}|>\sigma_{k}$, $k=1,\ldots, N$, and $\pi k_{i}/\sigma_{i}=\hat k_{i}$, then $f(t_{1} ,\ldots…,t_{N} )=$

\begin{displaymath}
\sum^{\infty}_{k_{1} =-\infty}\negthickspace\cdots\negthickspace\negthickspace\negthickspace \sum^{\infty}_{k_{N}=-\infty} \negthickspace\negthickspace f\left(\hat k_{1} ,.,\hat k_{N}\right)sinc( \sigma_{1}(t_{1}- –\hat k_{1}) ).. sinc(\sigma_{N}(t_{N}-\hat k_{N} ))
\end{displaymath}

Therefore by Parzen's theorem, on $\mathcal{I}_{0}$, $\Phi^{0}(\mathbf{z})=$

$$\sum_{n\in \mathcal{I}_{0}}\Gamma_{n}A^{3}T_{\mathbf{m}_{n'}}sinc_{t_{P},l_{P}}(\mathbf{z})=$$

$$\sum_{n\in \mathcal{I}_{0}}\Psi(\mathbf{m}_{n})A^{3}T_{\mathbf{m}_{n'}}sinc_{t_{P},l_{P}}(\mathbf{z})\approx \Psi(\mathbf{z})$$.

Assume that the process has generated all generations up to and including $m+1$. Let $g_{n'}(\mathbf{z})=A^{3}T_{\mathbf{m}_{n'}}sinc_{t_{P},l_{P}}(\mathbf{z})$. Recall that $\Phi^{m+1}(\mathbf{s})=\sum_{n\in \mathcal{I}_{m+1}}\negthickspace\Phi_{n}(\mathbf{z})$.

Hence one can write 

$$\Phi^{m+1}(\mathbf{z})=\sum_{n'\in \mathcal{I}_{m+1}}\sum_{n\in L_{n'}}\mathcal{T}_{n'n}g_{_{n^{m+1}}}(\mathbf{z})$$ 

If we assume the convention that $\mathcal{T}_{n'n}=0$ if $n$ does not propagate information to $n'$ then we can rewrite the above as 

$$\Phi^{m+1}(\mathbf{z})=\sum_{n^{m+1+}\in \mathcal{I}_{m+1}}\sum_{n^{m}\in \mathcal{I}_{m}}\mathcal{T}_{n^{m+1}n^{m}}g_{n^{m+1}}(\mathbf{z})$$ 
$$=\sum_{n^{m+1}\in \mathcal{I}_{m+1}}\sum_{n^{m}\in \mathcal{I}_{m}}\mathcal{P}_{n^{m+1}n^{m}}\phi_{n^{m}}g_{n^{m+1}}(\mathbf{z})$$

$$=\sum_{n^{m+1}\in \mathcal{I}_{m+1}}\sum_{n^{m}\in \mathcal{I}_{m}}\mathcal{P}_{n^{m+1}n^{m}}\negthickspace\negthickspace\negthickspace\sum_{n^{m-1}\in \mathcal{I}_{m-1}}\negthickspace\negthickspace\negthickspace\mathcal{T}_{n^{m}n^{m-1}}g_{n^{m+1}}(\mathbf{z})=$$

$$\negthickspace\sum_{n^{m+1}\in \mathcal{I}_{m+1}}\sum_{n^{m}\in \mathcal{I}_{m}}\mathcal{P}_{n^{m+1}n^{m}}\negthickspace\negthickspace\negthickspace\negthickspace\sum_{n^{m-1}\in \mathcal{I}_{m-1}}\negthickspace\negthickspace\negthickspace\negthickspace\mathcal{P}_{n^{m}n^{m-1}}\phi_{n^{m-1}}g_{n^{m+1}}(\mathbf{z})$$

Continuing one arrives at

\begin{widetext}
$$\Phi_{m+1}(\mathbf{z}) = \sum_{n^{m+1}\in \mathcal{I}_{m+1}}\cdots \sum_{n^{0}\in \mathcal{I}_{0}}\mathcal{P}_{n^{m+1}n^{m}}\cdots\mathcal{P}_{n^{1}n^{0}}\phi_{n{0}}g_{{n^{m+1}}}(\mathbf{z})=$$
$$\sum_{n^{m+1}\in \mathcal{I}_{m+1}}\cdots \sum_{n^{0}\in \mathcal{I}_{0}}\mathcal{P}_{n^{m+1}n^{m}}\cdots\mathcal{P}_{n^{1}n^{0}}\Psi(\mathbf{m}_{n^{0}})g_{n^{m+1}}(\mathbf{z})=$$
$$\sum_{n^{m+1}\in \mathcal{I}_{m+1}}\sum_{n^{m}\in \mathcal{I}_{m}}\cdots \sum_{n^{0}\in \mathcal{I}_{0}}\frac{l_{P}^{3}}{A^{3}}e^{\frac{i}{\hbar}\tilde S[n^{m+1},n^{m}]}\frac{l_{P}^{3}}{A^{3}}e^{\frac{i}{\hbar}\tilde S[n^{m},n^{m-1}]}\times \cdots\times$$
$$\frac{l_{P}^{3}}{A^{3}}e^{\frac{i}{\hbar}\tilde S[n^{1},n^{0}]}\Psi(\mathbf{m}_{n^{0}})g_{n^{m+1}}(\mathbf{z})=$$
$$\sum_{n^{m+1}\in \mathcal{I}_{m+1}}\sum_{n^{m}\in \mathcal{I}_{m}}\cdots \sum_{n^{0}\in \mathcal{I}_{0}}e^{\frac{i}{\hbar}\tilde S[n^{m+1},n^{m}]+\tilde S[n^{m},n^{m-1}]+\cdots +\tilde S[n^{1},n^{0}]}\times $$
$$\overbrace{\frac{l_{P}^{3}}{A^{3}}\frac{l_{P}^{3}}{A^{3}}\cdots \frac{l_{P}^{3}}{A^{3}}}^{m+1}\Psi(\mathbf{m}_{n^{0}})g_{n^{m+1}}(\mathbf{z})=$$
$$\sum_{n^{m+1}\in \mathcal{I}_{m+1}}\sum_{n^{m}\in \mathcal{I}_{m}}\cdots \sum_{n^{0}\in \mathcal{I}_{0}}e^{\frac{i}{\hbar}\mathcal{L}[n^{m+1},n^{m}]t_{P}+ \mathcal{L}[n^{m},n^{m-1}]t_{P}+\cdots + \mathcal{L}[n^{1},n^{0}]t_{P}}\times$$ $$\overbrace{\frac{l_{P}^{3}}{A^{3}}\frac{l_{P}^{3}}{A^{3}}\cdots \frac{l_{P}^{3}}{A^{3}}}^{m+1}\Psi(\mathbf{m}_{n^{0}})g_{n^{m+1}}(\mathbf{z})\approx$$

$$\sum_{n^{m+1}\in \mathcal{I}_{m+1}}\sum_{n^{m}\in \mathcal{I}_{m}}\cdots \sum_{n^{0}\in \mathcal{I}_{0}}e^{\frac{i}{\hbar}S[n^{m+1},n^{0}]}
\overbrace{\frac{l_{P}^{3}}{A^{3}}\frac{l_{P}^{3}}{A^{3}}\cdots \frac{l_{P}^{3}}{A^{3}}}^{m+1}\Psi(\mathbf{m}_{n^{0}})g_{n^{m+1}}(\mathbf{z})\approx$$
$$\sum_{n^{m+1}\in \mathcal{I}_{m+1}}\sum_{n^{0}\in \mathcal{I}_{0}}\int_{I_{i-1}}\cdots \int_{I_{1}}e^{\frac{i}{\hbar}S[n^{m+1},n^{0}]}\overbrace{\frac{dx^{m+1}}{A^{3}}\cdots \frac{dx^{1}}{A^{3}}}^{m}\frac{l_{P}^{3}}{A^{3}}\Psi(\mathbf{m}_{n^{0}})g_{n^{m+1}}(\mathbf{z})$$
\end{widetext}

\noindent where the $I_{i}$ refers to the continuous extension of the sub-lattice upon which $\mathcal{I}_{i}$ is defined, i.e. $I_{i}=\{it_{P}\}\times \mathbb{R}^{3}$, $dx^{k}$ is a  differential on $I_{i}$ and the action integral has been taken over the piecewise linear path $l_{n^{0}},l_{n^{1}},\ldots,l_{n^{m+1}}$ on the continuous lattice extension $[0,(m+1)t_{P}]\times \mathbb{R}^{3}$ and where the final step is obtained approximating each discrete sum by an integral.

Now as $N,r\rightarrow \infty$, the number of informons from $\mathcal{I}_{k}$ contributing to any informon of $\mathcal{I}_{k+1}$ grows to infinity for each $k$ and a moment's reflection will suggest therefore that in the limit every possible path between the informons of $\mathcal{I}_{0}$ and the informons of $\mathcal{I}_{m+1}$ will be included in the calculation. The entirety of each causal tapestry will be connected to all of the other causal tapestries. As $t_{P}\rightarrow 0$, the temporal spacing between lattice slices decreases, so not only does the total number of lattice slices increase, but the number of lattice slices between $l_{0}$ and $l_{i}$ increases while their distance decreases. Note that these functions and integrals may be transferred to the causal manifold $\mathcal{M}$ via the causal embedding. It is easy to see that under such circumstances, according to Feynman and Hibbs \cite{Feynman} the product of the integrals above converges to the path integral between the points $\alpha_{0}$ and $\alpha_{i}$, and hence between $\mathbf{m}_{i}$ and $\mathbf{m}_{0}$.  Noting this,  one obtains $\Phi^{m+1}(\mathbf{z})\approx$

\begin{multline*}
 \sum_{n^{m+1}\in \mathcal{I}_{m+1}}\sum_{n^{0}\in \mathcal{I}_{0}}K(l_{n^{m+1}},l_{n^{0}})\Psi(\mathbf{m}_{n^{0}})\frac{l_{P}^{3}}{A^{3}}g_{n^{m+1}}(\mathbf{z})=\\
\sum_{n^{m+1}\in \mathcal{I}_{m+1}}\sum_{n^{0}\in \mathcal{I}_{0}}K(\mathbf{m}_{n^{m+1}},\mathbf{m}_{n^{0}})\Psi(\mathbf{m}_{n^{0}})\frac{l_{P}^{3}}{A^{3}}g_{n^{m+1}}(\mathbf{z})=\\
\negthickspace\sum_{n^{m+1}\in \mathcal{I}_{m+1}}\sum_{n^{0}\in \mathcal{I}_{0}}K(\mathbf{m}_{n^{m+1}},\mathbf{m}_{n^{0}})\Psi(\mathbf{m}_{n^{0}})l_{P}^{3}T_{\mathbf{m}_{n^{m+1}}}sinc_{t_{P},l_{P}}(\mathbf{z})\\
\approx \sum_{n^{m+1}\in \mathcal{I}_{m+1}}\int_{I_{0}}K(\mathbf{m}_{n^{m+1}},\mathbf{m}_{n^{0}})\Psi(\mathbf{m}_{0})dx_{0}T_{\mathbf{m}_{n^{m+1}}}sinc_{t_{P}l_{P}}(\mathbf{z})=\\
\sum_{n^{m+1}\in \mathcal{I}_{m+1}}\Psi(\mathbf{m}_{n^{m+1}})T_{\mathbf{m}_{n^{m+1}}}sinc_{t_{P}l_{P}}(\mathbf{z})=\Psi(\mathbf{z})
\end{multline*}

\noindent where the final step again uses Parzen's theorem. Technically, the final equality holds on the $t=(m+1)t_{P}$ time slice and the relation holds approximately on that portion of $\mathcal{M}$ for which $t\leq (m+1)t_{P}$ but only weakly on the forward portion for which $t>(m+1)t_{P}$.

\section{Errors}

The next question to address is the goodness of fit between the wave function as determined by this model and the wave function as calculated using the usual path integral or Schr\"odinger equation methods. Goodness of fit is a more accurate term because the truly important question is not whether it accurately matches the NRQM wave function but rather how well it satisfies the Schr\"odinger equation and provides the essential statistical relations. Comparison to the NRQM wave function is made by using the discrepancy measure $\rho$ or by substituting into the appropriate Schr\"odinger equation and examining for goodness of fit. This in turn determines whether or not the game is a win for Player II. If it is, then we can say that the reality game  $\mathfrak{R}(N,r,\rho,\delta,t_{P},l_{P},\omega,L,\Sigma,p)$ generates the wave function to accuracy $\delta$.

The discrepancy between the global $\mathcal{H}(\mathcal{M})$-interpretation given above and the standard NRQM wave function depends upon the accuracy of the approximation to the integral $\int_{\mathcal{M}_{t}} K(\mathbf{x}_{j'},\mathbf{x}_{j})\phi_{j}(\mathbf{x}_{j}) d\mathbf{x}_{j}$, the deviations from uniformity of the causal embedding points, the number $r$ of current informons contributing information to any nascent informon as well as the values of $t_{P},l_{P}$.  This is a difficult problem to assess in general but results are available in special cases. For example, in one dimension, if the wave function $\Psi$ satisfies $|\Psi|\leq M|t|^{-\gamma}$ for $0<\gamma\leq 1$, $|\int_{\mathcal{M}_{t}} K(x_{j'},x_{j})\phi_{j}(x_{j}) dx_{j}-\Psi(x_{j'})|\leq \epsilon$, the discrepancy between each embedding point and its ideal lattice embedding point is less that $\delta$, and the truncation number $r=2[W^{1+1/\gamma}+1]+1$, then according to a theorem of Butzer \cite{Zayed}, the error $E$ satisfies

\begin{displaymath}
||E||_{\infty}\leq -K(\Psi,\gamma,\epsilon/l_{P},\delta/l_{P})l_{P}\ln l_{P}
\end{displaymath}

\noindent where

$$K=(1+\frac{1}{\gamma})\left\{\sqrt{5}e\left[ (\frac{14}{\pi}+\delta/l_{P}+\frac{7}{3\sqrt{5}\pi})||\Psi^{(1)}||_{\infty}+\epsilon/l_{P}\right]+\\
 6e(M+||\Psi||_{\infty}) \right\}$$

Hence, $||E||_{\infty}\approx 10^{-33}K$ if $l_{P}$ is the Planck length.

In an ideal case in which the kernel sum equals the kernel integral, the causal embedding is uniform, and each round takes $t_{P}$ seconds so that a complete game lasts $|c|t_{P}$ seconds (where $|c|$ means the numerical value of the speed of light without the units), one can use the Yao and Thomas theorem \cite{Zayed} to find that the error roughly $T$ satisfies

\begin{displaymath}
|\Phi(\mathbf{z})-\Psi(\mathbf{z})|=|T(\mathbf{z})|\leq \frac{8\max_{\mathcal{M}'} |\Psi(x,y,z)|}{(2\pi)^{3}|c|^{3}}\approx 10^{-27}\;(m^{-3})
\end{displaymath}

Of course in general the accuracy of the finite approximations to the kernel integrals will need to be taken into account but the above calculation shows that even a simplified model is capable of generating a highly accurate NRQM wave function. This supports the contention that the process approach provides a viable alternative formulation for NRQM. The discussion now turns to the process algebra and its applications to the paradoxes of quantum mechanics.

\section{Interactions and the Algebra of Process}

 The generation of informons can trigger the activation or inactivation of processes or couplings between processes depending upon their compatibility \cite{Trofimova1}. Couplings between processes may take many forms. They may be competitive, cooperative, correlated, anticorrelated or may take more subtle or complex forms. Since the possibilities for interactions between processes are legion, the problem of interaction will be discussed generically with the understanding that the details will need to be spelled out in each individual case.

The various types of couplings between processes gives rise to the process algebra.
These couplings are distinguished by the manner in which rounds are organized for the generation of informons. Note that at this level of discussion the information content of the informons is not being addressed and must be determined from the specification of the individual processes. Rounds occur sequentially. Primitive processes generate single informons per round. Suppose that one has two processes $\mathbb{P}_{1}$ and $\mathbb{P}_{2}$. In a single round, $\mathbb{P}_{1}$ and $\mathbb{P}_{2}$ may generate their informons concurrently or they may generate them sequentially. Couplings between processes that generate their informons sequentially will be denoted by \emph{sums}. Couplings between processes that generate informons concurrently will be denoted by \emph{products}. If the order of generation is fixed for all rounds then the sum is said to be ordered. If the order can vary from round to round then it will be said to be unordered. In this paper only unordered sums will be considered. Products, by definition, are only unordered. 

Processes may act completely independently of one another or the actions of one process may constrain the actions of another. The simplest distinction concerns whether or not two processes may contribute information to the same informon. If they do not then the coupling is said to be \emph{exclusive}. Otherwise the coupling is said to be \emph{free}. Exclusive couplings are somewhat analogous to interactions among fermions while free couplings are somewhat like bosonic interactions. Finally, apart from any free/exclusive considerations, processes may act completely independently of one another (\emph{independent}) or their actions may constrain the actions of other processes, resulting in correlations or anticorrelations among the informons generated by the coupled processes. In such a case the coupling is termed \emph{interactive}. The interactive coupling is actually shorthand for a wide variety of possible relationships among the coupled processes and needs to be expressed in each case. Sometimes it can be described in terms of the other sums and products but sometimes it can only be described by means of the sequence tree, which is described in the next section.  These consideration give rise to 8 general possibilities - a) Sequential sums: $\hat\oplus$ (free, independent), $\oplus$ (exclusive, independent), $\hat\boxplus$ (free, interactive), $\boxplus$ (exclusive, interactive), b) Concurrent products: $\hat\otimes$ (free, independent), $\otimes$ (exclusive, independent), $\hat\boxtimes$ (free, interactive), $\boxtimes$ (exclusive, interactive).
The process algebra is clearly much richer than the algebra associated with Hilbert spaces and their operators.

If $\mathbb{P}$ is a primitive process generating a single entity, then a process generating $N$ such independent entities could be given by either $\overbrace{\mathbb{P}\hat\otimes\cdots \hat\otimes\mathbb{P}}^{N}$  or $\overbrace{\mathbb{P}\otimes\cdots \otimes\mathbb{P}}^{N}$ (a more accurate approach will be discussed in a later section).

Many meanings can be ascribed to the multiplication of a process by a scalar $w$. Here $w\mathbb{P}$ shall mean that any informon $[n]<\mathbf{m}_{n},\phi_{n}(\mathbf{z}),\Gamma_{n},\mathbf{p}_{n}>\{G\}$ of the process $\mathbb{P}$ shall be replaced by the informon $[n]<\mathbf{m}_{n},w\phi_{n}(\mathbf{z}),w\Gamma_{n},\mathbf{p}_{n}>\{G\}$.   

The \emph{character} of a process refers to the formal mathematical types of its various properties (scalar, vector, tensor, real, complex, quaternion, octonion) as opposed to their numerical \emph{values}. A process is categorized by its character, while states of a process are distinguished by their values. In determining the choice of algebraic operation to represent a coupling a reasonable rule of thumb is that processes possessing the same character may form sums or products, but processes possessing different characters may only form products. Moreover, different states of the same process sum only through the exclusive sum which ensures that an informon corresponding to a single process may only possess properties associated with a single state. This excludes the possibility of a single informon simultaneously representing multiple states of a single process. In other words, superposition only occurs at the level of process. A single state may be decomposed into a sum of subprocesses, such as occurs in the presence of boundary conditions, in which case the free sum must be used because in such a case the subprocesses all possess the same character and values. The sub-division of a single state process via boundary conditions does  not alter any intrinsic properties of the process, thus each subprocess is merely a restricted action of the same single state process, and therefore it is perfectly acceptable for these different subprocesses to contribute to the generation of a single informon even if over the course of several rounds.

The zero process, $\mathbb{O}$, is the process that does nothing. It generates no informons at all. From the discussion above it is reasonable to assert that processes possessing different characters cannot sum, so for example a scalar process $\mathbb{P}_{s}$ and a spinorial process $\mathbb{P}_{sp}$ would form the sum $\mathbb{P}_{s}\oplus \mathbb{P}_{sp}=\mathbb{O}$. The same might also hold for certain products of processes.. 

Processes may also be concatenated. Given a sequence of active processes 
\begin{displaymath}
\mathcal{I}_{n}\stackrel{\mathbb{P}_{n}}{\rightarrow}\mathcal{I}_{n+1}\stackrel{\mathbb{P}_{n+1}}{\rightarrow}\mathcal{I}_{n+2}
\end{displaymath}

one may form the concatenated process

\begin{displaymath}
\mathcal{I}_{n}\stackrel{\mathbb{P}_{n}\mathbb{P}_{n+1}}{\rightarrow}\mathcal{I}_{n+2}
\end{displaymath}

In general concatenations are non-commutative.

\section{Process Covering Map}
 
In the process model, a complete generation of informons represents an instance of reality and this reality unfolds through the successive actions of processes

\begin{displaymath}
\mathcal{I}\cdots\mathcal{I}_{n}\stackrel{\mathbb{P}_{n}}{\rightarrow}\mathcal{I}_{n+1}\stackrel{\mathbb{P}_{n+1}}{\rightarrow}\mathcal{I}_{n+2}\cdots
\end{displaymath}

Each generation $\mathcal{I}_{n}$ consists of a discrete and at most countable collection of informons and the information required by process $\mathbb{P}_{n}$ to generate $\mathcal{I}_{n+1}$ from $\mathcal{I}_{n}$ resides \emph{wholly} within the informons of $\mathcal{I}_{n}$ and relies \emph{only} on their intrinsic features. Thus the dynamics in the process model is consistent and self contained. The link to quantum mechanics is through the extrinsic features, the interpretations, which link informons to spacetime through their embedding to some causal manifold $\mathcal{M}$ and to physical entities through their association with elements of the Hilbert space $\mathcal{H}(\mathcal{M})$ from which global wave functions emerge through an interpolation procedure. 
The link between the process dynamics and quantum mechanics is given by the Process Covering Map (PCM). This map will first be constructed on the space of primitive processes, $\Pi_{p}$.

Let the prior generation of informons be the causal tapestry $\mathcal{I}_{n}$ and consider the action of some primitive process $\mathbb{P}$ in constructing $\mathcal{I}_{n+1}$. Once activated, $\mathbb{P}$ will generate a sequence of $n_{1}, n_{2}, \ldots, n_{k}, \ldots$ of informons, one informon corresponding to each round. Due to the non-deterministic nature of the process action, if one imagined reactivating $\mathbb{P}$ on $\mathcal{I}_{n}$ then a different sequence of informons, say $m_{1}, m_{2}, \ldots, m_{k}, \ldots$, would likely result. The sequence tree of a process $\mathbb{P}$ and initial generation $\mathcal{I}_{n}$, denoted $\Sigma(\mathbb{P},\mathcal{I}_{n})$ is a directed graph which keeps track of these various possible histories and is constructed as follows. Into level $0$ put the empty causal tapestry $\emptyset$. Into level $1$, place  all possible causal tapestries that can be generated by $\mathbb{P}$ in the first round starting from $\mathcal{I}_{n}$. The edge between $\emptyset$ and one such tapestry $\mathcal{I}^{1}_{k}=\{n\}$ is just the informon $n$ generated in the round. Level 2 is formed by completing the next round beginning with $\mathcal{I}_{n}$ and some $\mathcal{I}^{1}_{k}$. If some informon $m$ is generated then one adds an edge $m$ connecting $\mathcal{I}^{1}_{k}$ to $\mathcal{I}^{2}_{j}=\mathcal{I}^{1}_{k}\cup \{m\}$.\  One continues in this manner until the action of $\mathbb{P}$ is complete. A 
path along the sequence tree from $\emptyset$ to some final causal tapestry ${}_{j}\mathcal{I}_{n+1}$ represents one possible outcome of $\mathbb{P}$.

If ${}_{j}\mathcal{I}_{n+1}=\{n_{1}, n_{2}, \ldots, n_{k}, \ldots\}$ then a global $\mathcal{H}(\mathcal{M})$ interpretation ${}_{j}\Phi_{n+1}(\mathbf{z})$ can be associated with ${}_{j}\mathcal{I}_{n+1}$ by defining ${}_{j}\Phi_{n+1}(\mathbf{z})=\sum_{n_{i}\in {}_{j}\mathcal{I}_{n+1}}\phi_{n_{i}}(\mathbf{z})$ where $\phi_{n_{i}}(\mathbf{z})$ is the local $\mathcal{H}(\mathcal{M})$ interpretation of the informon $n_{i}$.
Thus to the sequence tree  $\Sigma(\mathbb{P},\mathcal{I}_{n})$ one may associate a \emph{set} of global $\mathcal{H}(\mathcal{M})$ interpretations $H(\mathbb{P},\mathcal{I}_{n+1})=\{{}_{1}\Phi_{n+1}(\mathbf{z}),\ldots,{}_{k}\Phi_{n+1}(\mathbf{z}),\ldots\}$, one corresponding to each path down the tree. The PCM is defined as follows: for a primitive process $\mathbb{P}$ and initial causal tapestry $\mathcal{I}$ set $\mathfrak{P}(\mathbb{P},\mathcal{I})=H(\mathbb{P},\mathcal{I})$. If $\mathcal{I}$ is fixed or unspecified one can write $\mathfrak{P}_{\mathcal{I}}(\mathbb{P})$ or just $\mathfrak{P}(\mathbb{P})$.
The PCM is thus a set valued map \cite{Aubin}. This is a further distinction from quantum mechanics.

Interpolation theory \cite{Maymon,Zayed,Marks,Sulisthesis} shows that given certain choices of the interpolation function $g$, in the limit $N,r\rightarrow \infty$, $\mathfrak{P}(\mathbb{P},\mathcal{I})\rightarrow \{\Phi^{t_{P}l_{P}}(\mathbf{z})\}$, a single function, where $t_{P},l_{P}$ are interpolation parameters. $N\rightarrow\infty$ means that the generators cover the entirety of a spacelike hypersurface of the causal manifold and $r\rightarrow\infty$ means that the amount of information transferred from the prior causal tapestry to each nascent informon is complete. As was illustrated in a Section IV, it is in this limit that (non-relativistic) quantum mechanics emerges.

It is possible, given two distinct primitive processes $\mathbb{P}_{1}$ and $\mathbb{P}_{2}$, that in the limit  $N,r\rightarrow \infty$, $\mathfrak{P}(\mathbb{P}_{1},\mathcal{I})=\mathfrak{P}(\mathbb{P}_{2},\mathcal{I})$. In such a case the primitive processes $\mathbb{P}_{1},\mathbb{P}_{2}$ are said to be $\Psi$-\emph{epistemic equivalent} (alternatively \emph{weak epistemic equivalent}). $\Psi$-epistemic equivalent processes generate distinct \textquotedblleft realities\textquotedblright\, at small scales yet asymptotically yield the same global interpretation. In the case that this global interpretation corresponds to a quantum mechanical wave function, these two processes would be indistinguishable from the standpoint of quantum mechanics. 

Strictly speaking, the current state of the art in interpolation theory applies primarily to vector valued functions, so these comments are limited to a discussion of integer spin, and most especially scalar, particles. Nevertheless there is no theoretical reason why this should not also extend to the case of half integer spin particles as well.

The PCM is now extended to the space of general processes, $\Pi$, by considering sums and products. 
First consider an independent exclusive sum $\oplus_{i} w_{i}\mathbb{P}_{i}$ of primitive processes $\mathbb{P}_{i}$ acting on the causal tapestry $\mathcal{I}$. Since this is an exclusive sum, the informons of the nascent causal tapestry  $\mathcal{I}_{1}$ will lie in distinct subsets, $\mathcal{I}_{1}^{i}$, each corresponding to a specific subprocess $\mathbb{P}_{i}$. Let $j_{n}=i$ iff $n\in \mathcal{I}_{1}^{i}$. Since this is an independent sum, the actions of each subprocesses can be considered independently of the others. Then the global $\mathcal{H}(\mathcal{M})$-interpretation $\Phi(\mathbf{z})=\sum_{n\in \mathcal{I}_{1}} w_{j_{n}}\phi_{n}(\mathbf{z})=\sum_{i}w_{i}\{\sum_{n\in \mathcal{I}_{1}^{i}}\phi_{n}(\mathbf{z})\}=\sum_{i}w_{i}\Phi^{i}(\mathbf{z})$ where $\Phi^{i}(\mathbf{z})$ is the global $\mathcal{H}(\mathcal{M})$-interpretation corresponding to the process $\mathbb{P}_{i}$. Applying this to each path of the sequence tree it follows easily that $$\mathfrak{P}(\oplus_{i} w_{i}\mathbb{P}_{i})=\sum_{i} w_{i}\mathfrak{P}(\mathbb{P}_{i})$$ where for two sets of functions $A,B$ the sum $A+B=\{f+g|f\in A,g\in B\}$.

It is not difficult to show that this also holds true for free sums, but in this case these subsets may overlap and informons must be artificially divided to reflect the contributions from each subprocess. Nevertheless,$$\mathfrak{P}(\hat\oplus_{i} w_{i}\mathbb{P}_{i})=\sum_{i} w_{i}\mathfrak{P}(\mathbb{P}_{i})$$.

Here the process approach clearly departs  from  quantum mechanics. Obviously $\oplus_{i} w_{i}\mathbb{P}_{i}$ and $\hat\oplus_{i} w_{i}\mathbb{P}_{i}$ are $\Psi$-epistemic equivalent, but the processes $\oplus_{i} w_{i}\mathbb{P}_{i}$ and $\hat\oplus_{i} w_{i}\mathbb{P}_{i}$ possess very different interpretations on the causal manifold $\mathcal{M}$ and may possess quite distinct sequence trees. The local $\mathcal{H}(\mathcal{M}))$-interpretations will also differ even though the asymptotic global interpretations are the same.

It is not possible, in general, to describe interactive sums, which must be determined from their sequence trees.   

The case of products is more complicated. First consider an independent exclusive product $\otimes_{i} \mathbb{P}_{i}$ of primitive processes $\mathbb{P}_{i}$. During each round  a set of informons $\{n^{i}\}$ will generated together with a set of their local Hilbert space contributions, $\{\phi_{n^{i}}(\mathbf{z})\}$.  The most natural representation is to consider the co-product of the Hilbert spaces corresponding to each subprocess. This maintains the point of view of individual entities. The usual approach in quantum mechanics, however, is to consider the product of the Hilbert spaces. In either case each edge in the sequence tree may be given by a tuple $(n^{1},n^{2},\ldots,n^{j})$ of informons. There will be corresponding tuples of causal manifold points $(\mathbf{m}_{n^{1}},\mathbf{m}_{n^{2}},\ldots,\mathbf{m}_{n^{j}})$ and of local Hilbert space contributions $(\phi_{n^{1}}(\mathbf{z}),\phi_{n^{2}}(\mathbf{z}),\ldots,\phi_{n^{j}}(\mathbf{z}))$. The vertices of the sequence tree may be recursively defined in the co-product case $\mathcal{I}_{n}\rightarrow \mathcal{I}_{n}\cup_{i} \{n^{i}_{n}\}$ and in the product case as $\mathcal{I}^{1}_{n}\times\cdots\times\mathcal{I}^{n}_{n}\rightarrow (\mathcal{I}^{1}_{n}\cup \{n^{1}_{n}\})\times\cdots\times(\mathcal{I}^{n}_{n}\cup\{n^{n}_{n}\})$.  Since this product is independent exclusive it follows easily that 
 
 $$\mathfrak{P}(\otimes_{i} \mathbb{P}_{i})=\mathfrak{P}(\mathbb{P}_{1})\oplus\mathfrak{P}(\mathbb{P}_{2})\oplus\cdots\oplus\mathfrak{P}(\mathbb{P}_{j})$$
 
\noindent in the co-product case (where $\oplus$ means a formal sum of functions, not a pointwise sum) and

 $$\mathfrak{P}(\otimes_{i} \mathbb{P}_{i})=\{(\Phi_{n^{1}}(\mathbf{z}),\Phi_{n^{2}}(\mathbf{z}),\ldots,\Phi_{n^{j}}(\mathbf{z}))|\text{ over all instances of } \mathbb{P}\}=$$
$$\mathfrak{P}(\mathbb{P}_{1})\times\mathfrak{P}(\mathbb{P}_{2})\times\cdots\times\mathfrak{P}(\mathbb{P}_{j})$$
\noindent in the product case, where $\times$ means a set product, not a pointwise product.

It can also be shown that, as for sums, the free product satisfies

$$\mathfrak{P}(\hat\otimes_{i} \mathbb{P}_{i})=\mathfrak{P}(\mathbb{P}_{1})\times\mathfrak{P}(\mathbb{P}_{2})\times\cdots\times\mathfrak{P}(\mathbb{P}_{j}) \;(or  \;\mathfrak{P}(\mathbb{P}_{1})\oplus\mathfrak{P}(\mathbb{P}_{2})\oplus\cdots\oplus\mathfrak{P}(\mathbb{P}_{j}))$$

\noindent so that $\otimes_{i} \mathbb{P}_{i}$ and $\hat\otimes_{i} \mathbb{P}_{i}$ are also $\Psi$-epistemic equivalent. This again shows the incompleteness of the quantum mechanical framework.

The situation for interactive products is much more complicated. In some cases it may be possible to express it in terms of independent sums and products but in general the PCM will need to be derived directly from an examination of the sequence tree.

The PCM shows that the relative strength of a process $\mathbb{P}$ is not an invariant but instead depends upon the particular causal tapestry that has been generated from the initial causal tapestry during an application of the process. In the asymptotic limit $N,r\rightarrow\infty$, the PCM becomes single valued, yielding the Hilbert space interpretation $\Phi^{t_{P}l_{P}}(\mathbf{z})$. For fixed values of $t_{P},l_{P}$, this function is fixed and so one may define the strength of $\mathbb{P}$, $||\mathbb{P}||$ as $||\mathbb{P}||^{2}=\sum_{n_{{i}}\in \mathcal{I}_{\infty}} l^{3}_{P}\Gamma^{*}_{n_{i}}\Gamma_{n_{i}}$. In the case of sinc interpolation one can show that $$\sum_{n_{{i}}\in \mathcal{I}_{\infty}} l^{3}_{P}\Gamma^{*}_{n_{i}}\Gamma_{n_{i}}= ||\Phi(\mathbf{z})^{t_{P}l_{P}}||^{2}=\negmedspace
\int_{M}\Phi^{*}(\mathbf{z})\Phi(\mathbf{z})=\negmedspace
\int_{M}\sum_{n_{i}}\sum_{n_{j}}\Phi^{t_{P}l_{P}*}_{n_{i}}(\mathbf{z})\Phi^{t_{P}l_{P}}_{n_{j}}(\mathbf{z})$$ \cite{Zayed}. Hence $||\mathbb{P}||^{2}=||\Phi(\mathbf{z})^{t_{P}l_{P}}||^{2}$ and this process strength is well defined.

\section{Process Approach to the Configuration Space}

The PCM defined in the previous section essentially provides a spacetime representation, especially in the case that the co-product representation is used for process products. Even if the product representation is used, the PCM still provides a description of individual subprocesses, even if they are formed into products. This corresponds to a picture of reality as consisting of the behaviour of individual physical entities and that is what is commonly observed. In quantum mechanics, however, one is most often interested in correlations between physical entities. The multi-particle wave function is frequently taken to be a function theoretic product (not a set theoretic product) of individual wave functions. These wave functions are usually defined on spacetime. However, in more complex situations the wave function is not defined on spacetime but rather on a configuration space, which is an abstract space representing different configurations of the multiple particles. There has been much debate about the reality of the configuration space.

The process theory of measurement postulates that measurement occurs as an interaction between some process of interest and a specialized process termed a measurement process \cite{Sulis2,Sulisjmp,Sulisthesis}. Different outcomes of this interaction are determined by local couplings of the system and measurement processes which in turn are triggered by the generation of particular informons which determine the likelihood of such couplings. Thus in order to study measurement from a process perspective it is helpful to keep track of which informons are generated during a given round. 

 The problem for forming a product representation lies in how the global $\mathcal{H}(\mathcal{M})$ interpretations are generated. During each round a product process $\otimes_{i=1}^{n}\mathbb{P}_{i}$ will generate a correlated set of informons $A_{i}=(n^{1}_{i},\ldots,n^{n}_{i})$. If $\mathcal{I}'$ is the causal tapestry consisting of these tuples and formed by a complete action of $\otimes_{i=1}^{n}\mathbb{P}_{i}$ then the global $\mathcal{H}(\mathcal{M})$ interpretation for the product process on $\mathcal{M}^{n}$ should reasonably be defined as 

$$\hat\Psi_{p}(\mathbf{z}_{1},\ldots,\mathbf{z}_{n})=\sum_{(n^{1}_{i},\ldots,n^{n}_{i})\in \mathcal{I}'}\Gamma_{n^{1}_{i}}\cdots\Gamma_{n^{n}_{i}}T_{\mathbf{m}_{n^{1}_{i}}}g(\mathbf{z}_{1})\cdots T_{\mathbf{m}_{n^{n}_{i}}}g(\mathbf{z}_{n})$$

\noindent which is consistent with the formulation of the global $\mathcal{H}(\mathcal{M})$ interpretation for primitive processes and where $g$ is the local interpretation function.

Unfortunately $\hat\Psi_{p}$ is inadequate for determining correlations. The problem with $\hat\Psi_{p}$ is that it is based upon a single complete action of the product process which  cannot take into account the effects of all possible actions of the process. The proper way to generate a function expressing these correlations is through  the configuration sequence tree. This generalizes the construction of the sequence tree described in the previous section.  For $n$ subprocesses, each vertex of the sequence tree will be a causal tapestry $\mathcal{K}_{j}$ consisting of a set of ordered tuples of $n$ informons of the form $(n^{1}_{i},\ldots,n^{n}_{i})$ which is just a tuple formed from the informons generated at round $i$ by the generating process. An edge will consist of an $n$-tuple of informons $(n^{1}_{j},\ldots,n^{n}_{j})$ so that if this edge takes $\mathcal{K}_{i}\rightarrow \mathcal{K}_{i+1}$ then $\mathcal{K}_{i+1}=\mathcal{K}_{i}\cup\{(n^{1}_{j},\ldots,n^{n}_{j})\}$. Let ${}_{i}\mathcal{K}$ denote a causal tapestry formed by completely traversing a path in the tree. Note that each informon generated by a subprocess $\mathbb{P}_{i}$ in the sequence tree is generated independently from all of the others generated by $\mathbb{P}_{i}$ and each edge set is generated independently from every other edge set. The causal tapestry ${}_{i}\mathcal{K}$ may be artificially extended by adding informons from ${}_{j}\mathcal{K}$ so long as  we ensure that if we wish to add an informon $(n_{j}^{1},\ldots ,n_{j}^{n})\in {}_{j}\mathcal{K}$  then it is necessary that for each component informon $n_{j}^{k}$ such that there exists an informon $(n_{g}^{1},\ldots,n_{g}^{n})\in {}_{i}\mathcal{K}$ with $\mathbf{p}_{n_{j}^{k}}=\mathbf{p}_{n_{g}^{k}}$ and  $\mathbf{m}_{n_{j}^{k}}=\mathbf{m}_{n_{g}^{k}}$ , then $\Gamma_{n_{j}^{k}}=\Gamma_{n_{g}^{k}}$. That is, if we form the projection ${}_{i}\mathcal{K}\rightarrow {}_{i}\mathcal{K}_{k}$ by mapping each informon $(n_{j}^{1},\ldots, n_{j}^{n})\rightarrow n_{j}^{k}$, then ${}_{i}\mathcal{K}_{k}$ forms a consistent causal tapestry in its own right. An informon of  ${}_{j}\mathcal{K}$ which meets this condition is  said to be \emph{admissible} for ${}_{i}\mathcal{K.}$  Define the \emph{consistent union} ${}_{i}\mathcal{K}\bigtriangleup{}_{j}\mathcal{K}$ to be the set ${}_{i}\mathcal{K}\cup\{n\in {}_{j}\mathcal{K}\text{ admissible in }{}_{i}\mathcal{K}\}$.  A causal tapestry $\mathcal{K}$ is said to be maximal for a sequence tree if there is no path and no causal tapestry $\mathcal{K}'$ generated by this path such that $\mathcal{K}\bigtriangleup \mathcal{K}'\neq\mathcal{K}$.

The configuration sequence tree is denoted $\Sigma^{C}(\mathcal{K},\otimes_{i} \mathbb{P}_{i})$. A similar construction holds for the free product as well and will be denoted $\Sigma^{C}(\mathcal{I}^{n},\hat\otimes_{i}\mathbb{P}_{i})$.

Given a configuration sequence tree $\Sigma^{C}(\mathcal{K},\mathbb{P})$ let $\mathfrak{I}^{M}_{\Sigma^{C}(\mathcal{K},\mathbb{P})}$ denote the set of all of its maximal causal tapestries. We define the configuration process covering map or PCM${}^{C}$, denoted $\mathfrak{P}^{C}(\otimes_{i} \mathbb{P}_{i},\mathcal{K})$ ( or sometimes  $\mathfrak{P}^{C}_{\mathcal{K}}(\otimes_{i} \mathbb{P}_{i})$ or $\mathfrak{P}^{C}(\otimes_{i} \mathbb{P}_{i})$)  to be $\mathfrak{P}^{C}_{\mathcal{K}}(\otimes_{i} \mathbb{P}_{i})=$

$$\{\Phi^{j}(\mathbf{z})=\sum_{(n^{1}_{k},\ldots,n^{n}_{k})\in \mathcal{K}'} \Gamma_{n^{1}_{k}}\cdots\Gamma_{n^{n}_{k}}T_{\mathbf{m}_{n^{1}_{k}}}g(\mathbf{z}_{1})\times\cdots \times T_{\mathbf{m}_{n^{n}_{k}}}g(\mathbf{z}_{n})|\mathcal{K}'\in\mathfrak{J}^{M}_{\Sigma^{C}(\mathcal{K},\otimes_{i}\mathbb{P}_{i})}\}$$

It may be the case that the maximal causal tapestries are full product causal tapestries. In such a case, the asymptotic limit takes the form 

$$\mathfrak{P}^{C}_{\mathcal{I}^{n}}(\otimes_{i}\mathbb{P}_{i})=\{\Psi_{t_{P}l_{P}}^{1}(\mathbf{z}_{1})\times \cdots \times \Psi_{t_{P}l_{P}}^{n}(\mathbf{z}_{n})\}$$

\noindent which is identical to the usual quantum mechanical wave function form for independent particles.

The situation for interactive products is much more complicated and it may not always be possible to find a closed form solution. Nevertheless, a global $\mathcal{H}(\mathcal{M})$ interpretation will \emph{always} exist for the process model. Again one sees that the process model is more general than quantum mechanics and provides more dynamical information while yielding the same statistical information. 
 
 The configuration space wave function arises as an asymptotic limit of the PCM${}^{C}$ so that in the process model the configuration space is understood as  merely an epistemological entity and not an ontological one. The PCM on the other hand has both epistemological and ontological aspects.
\section{Process and Operators}

The study of the relationship between processes and operators on a Hilbert space is unexplored and promising. Only a few remarks will be offered here.

In the discussion above, the PCM was referenced to a given causal tapestry $\mathcal{I}$ which served as an initial condition upon which the process $\mathbb{P}$ was to act. It is certainly possible that the initial condition could be the empty tapestry but in general it will represent the outcome of the actions of previous processes.

Recall that a causal tapestry $\mathcal{I}$ consists of  informons, each of which has the form $[n]<(\mathbf{m}_{n},\phi_{n}(\mathbf{z}),\Gamma_{n},\mathbf{p}_{n})>\{G_{n}\}$. The function $\phi_{n}(\mathbf{z})$ provides a local contribution from $n$ to a global function $\Phi_{\mathcal{I}}(\mathbf{z})$ defined on the causal manifold $\mathcal{M}$ by $\Phi_{\mathcal{I}}(\mathbf{z})=\sum_{n\in \mathcal{I}} \phi_{n}(\mathbf{z})$. This defines a mapping $\mathfrak{T}$ from the space of causal tapestries $\mathbf{I}$ to the Hilbert space $\mathcal{H}(\mathcal{M})$ by

$$\mathfrak{T}(\mathcal{I})=\Phi_{\mathcal{I}}(\mathbf{z})$$.

The PCM was defined as a map $\mathfrak{P}:\Pi\times \text{\textbf{I}}\rightarrow \mathcal{P}(\mathcal{H}(\mathcal{M}))$ (the power set on $\mathcal{H}(\mathcal{M})$). If we fix some process $\mathbb{P}\in \Pi$ then we can define a tapestry covering map (TCM) $\mathfrak{P}_{\mathbb{P}}:\text{\textbf{I}} \rightarrow \mathcal{P}(\mathcal{H}(\mathcal{M}))$ in the obvious manner.

Define a generalized operator $\mathcal{G}$ on $\mathcal{H}(\mathcal{M})$ as a mapping $\mathcal{G}:\mathcal{H}(\mathcal{M})\rightarrow \mathcal{P}(\mathcal{H}(\mathcal{M}))$ such that $\mathcal{G}(f+g)\subset\mathcal{G}(f)+\mathcal{G}(g)$ and let $\mathfrak{G}(\mathcal{H}(\mathcal{M}))$ denote the set of generalized operators on $\mathcal{H}(\mathcal{M})$.

For a fixed process $\mathbb{P}$, define a generalized operator $\mathfrak{G}_{\mathbb{P}}$ on $\mathcal{H}(\mathcal{M})$ such that for every $f\in \mathcal{H}(\mathcal{M})$,  $\mathfrak{G}_{\mathbb{P}}(f)=\cup_{\mathcal{I}\in \mathfrak{T}^{-1}(f)} \mathfrak{P}_{\mathbb{P}}(\mathcal{I}).$

One thus obtains the following diagram

\begin{displaymath}
\begin{array}{ccc}
\mathbf{I} & \stackrel{\mathfrak{P}_{\mathbb{P}}}{\rightarrow} & \mathcal{P}(\mathcal{H}(\mathcal{M})) \\
\mathfrak{T}\downarrow &  & \downarrow e \\
\mathcal{H}(\mathcal{M}) & \stackrel{\rightarrow}{\mathfrak{G}_{\mathbb{P}}} & \mathcal{P}(\mathcal{H}(\mathcal{M})) \\
\end{array}
\end{displaymath} 

\noindent where $e$ is a map such that $h\subset e(h)$.

The problem is that one cannot guarantee that if two causal tapestries $\mathcal{I},\mathcal{I}'$ satisfy $\mathfrak{T}(\mathcal{I})=\mathfrak{T}(\mathcal{I}')$ (that is, they generate the same global Hilbert space interpretation), then $\mathfrak{P}_{\mathbb{P}}(\mathcal{I})=\mathfrak{P}_{\mathbb{P}}(\mathcal{I}')$ (that is, the process $\mathbb{P}$ generates the same collection of global Hilbert space interpretations). A process $\mathbb{P}$ is said to be $\Psi$-\emph{faithful} if $\mathfrak{T}(\mathcal{I})=\mathfrak{T}(\mathcal{I}')$ implies that $\mathfrak{P}_{\mathbb{P}}(\mathcal{I})=\mathfrak{P}_{\mathbb{P}}(\mathcal{I}')$ for all $\mathcal{I},\mathcal{I}'$. In the case of a $\Psi$-faithful process the diagram reduces to the simpler form

\begin{displaymath}
\begin{array}{ccc}
\mathbf{I} & \stackrel{\mathfrak{P}_{\mathbb{P}}}{\rightarrow} & \mathcal{P}(\mathcal{H}(\mathcal{M})) \\
\mathfrak{T}\downarrow &  & \downarrow id \\
\mathcal{H}(\mathcal{M}) & \stackrel{\rightarrow}{\mathfrak{G}_{\mathbb{P}}} & \mathcal{P}(\mathcal{H}(\mathcal{M})) \\
\end{array}
\end{displaymath} 

\noindent where $id$ is the identity.

In either case one can associate each process $\mathbb{P}$ with a generalized operator $\mathfrak{G}_{\mathbb{P}}$ on $\mathcal{H}(\mathcal{M})$.

If we assume that the process $\mathbb{P}$ involves an effective interpolation strategy, then one can show \cite{Zayed} that in the limit as $N,r\rightarrow \infty$, each PCM becomes effectively a map from $\Pi$ to $\mathcal{H}(\mathcal{M})$ (or equally, each TCM becomes a map from $\mathbf{I}\rightarrow \mathcal{H}(\mathcal{M})$) since the outcome set is a singleton. Consider now the situation in which the asymptotic limit has been taken. This corresponds to restricting attention to only those processes corresponding to the asymptote, so those processes for which $N,r=\aleph_{0}$ (at least).  If we restrict then to the subset $\Pi_{\infty}$ of such asymptotic processes then one must also restrict the space of causal tapestries to $\mathbf{I}_{\infty}$, the subset of causal tapestries that are generated by processes within $\Pi_{\infty}$. 
Hence for $\mathbb{P}\in\Pi_{\infty}$ and $\mathcal{I}_{\infty}\in \mathbf{I}_{\infty}$,

$$\mathfrak{P}_{\mathcal{I}_{\infty}}(\mathbb{P})=\{\Phi^{t_{P}l_{P}}(\mathbf{z})\},$$ a singleton set. 
 
It follows that the previous diagrams simplify to

\begin{displaymath}
\begin{array}{ccc}
\mathbf{I}_{\infty} & \stackrel{\mathfrak{P}_{\mathbb{P}}}{\rightarrow} & \mathcal{H}(\mathcal{M}) \\
\mathfrak{T}\downarrow &  & \downarrow e \\
\mathcal{H}(\mathcal{M}) & \stackrel{\rightarrow}{\mathfrak{G}_{\mathbb{P}}} & \mathcal{P}(\mathcal{H}(\mathcal{M})) \\
\end{array}
\end{displaymath} 

for general processes and for $\Psi$-faithful processes to

\begin{displaymath}
\begin{array}{ccc}
\mathbf{I}_{\infty} & \stackrel{\mathfrak{P}_{\mathbb{P}}}{\rightarrow} & \mathcal{H}(\mathcal{M}) \\
\mathfrak{T}\downarrow &  & \downarrow id \\
\mathcal{H}(\mathcal{M}) & \stackrel{\rightarrow}{\mathfrak{G}_{\mathbb{P}}} & \mathcal{H}(\mathcal{M}) \\
\end{array}
\end{displaymath} 
  
In this situation, the generalized operator $\mathfrak{G}_{\mathbb{P}}$ becomes a standard operator on $\mathcal{H}(\mathcal{M})$. This is the main value for considering $\Psi$-faithful processes.

The association between processes and operators on $\mathcal{H}(\mathcal{M})$ strongly suggests that symmetry structure such as expressed in Lie groups and algebras should also appear in the process algebra. This idea is not developed here, but it is in accord with the convention that processes are generators of informons and of properties, and so symmetries involving such properties should be reflected in the algebraic structure of processes and inherited secondarily by informons. This in turn suggests that the symmetry structure of spacetime events, expressed by informons, is emergent, not intrinsic.

\section{Process Approach  to the Paradoxes}

Quantum mechanics abounds with paradoxes \cite{Greenstein}, conflicts with special and general relativity \cite{Maudlin1}, is plagued by divergences \cite{Folland} and raises deep conceptual problems in relation to our understanding of the nature of fundamental reality \cite{Aharonovp}. Nevertheless, it has proven to be one of physics most successful theories, both in terms of computations as well as predictions of new phenomena. As a theory concerning the statistical structure of fundamental reality it is without peers. Clearly there is something that is correct about quantum mechanics but at the same time there is also something not quite right about it. The process perspective suggests that what is wrong with quantum mechanics is its inability to depict in a realist manner the underlying dynamical behaviour of systems and to express subtle distinctions in the couplings of systems to one another. This lack of expressiveness leads to conceptual confusions. It is the ingredient necessary for the completion of quantum mechanics.

The discussion of the PCM, PCM${}^{C}$ and $\Psi$-epistemic equivalence all demonstrate that the process algebra possesses more dynamical information than does the Hilbert space algebra of quantum mechanics, which arises from the process algebra as an asymptotic limit and as a quotient. This is even more apparent when considering how the wave function arises in the process algebra.

Assume that some process $\mathbb{P}$ has generated a causal tapestry $\mathcal{I}$ consisting of the informons $\{n_{1},\ldots,n_{k},\ldots\}$. The global $\mathcal{H}(\mathcal{M})$ interpretation is given as

$$\Phi(\mathbf{z})=\sum_{n_{i}}\Gamma_{n_{i}}T_{\mathbf{m}_{n_{i}}}g(\mathbf{z})$$

\noindent With a suitable choice of process action and interpolation procedure, the global $\mathcal{H}(\mathcal{M})$ interpretation provides an approximation to some quantum mechanical wave function $\Psi(\mathbf{z})$ and the expression 

$$\Phi(\mathbf{z})=\sum_{n_{i}}\Gamma_{n_{i}}T_{\mathbf{m}_{n_{i}}}g(\mathbf{z})\approx\Psi(\mathbf{z})$$

\noindent becomes exact asymptotically.
Note, however, that $\Phi(\mathbf{z})$ is actually generated only on the causal manifold elements $\{\mathbf{m}_{n_{1}},\ldots,\mathbf{m}_{n_{k}},\ldots\}$ and not on all of $\mathcal{M}$. This is because the actual physical space is the causal tapestry $\mathcal{I}$, which is both finite and discrete. The use of $\mathcal{M}$ is an artifice, an interpretation constructed in the mind of some observer which connects observations to some physical theory. The global $\mathcal{H}(\mathcal{M})$ interpretation is an interpolation on $\mathcal{M}$ whose values at elements away from the generating elements must be calculated using the interpolation formula. They do not actually exist but are again an artifice, an interpretation which connects observation to theory. The resulting quantum mechanical wave function also becomes an interpretation which is incapable of providing any information about the generating set that supports it. Indeed interpolation theory (especially non-uniform interpolation theory) \cite{Zayed} shows that there are an infinite number of subsets $\{\mathbf{m}_{n'_{1}},\ldots,\mathbf{m}_{n'_{k}},\ldots\}$ of $\mathcal{M}$ which can generate the \emph{same} wave function $\Psi(\mathbf{z})$,

$$\Psi(\mathbf{z})=\sum_{n_{i}}\Gamma_{n_{i}}T_{\mathbf{m}_{n_{i}}}g(\mathbf{z})=\sum_{n'_{i}}\Gamma_{n'_{i}}T_{\mathbf{m}_{n'_{i}}}g(\mathbf{z})$$

\noindent so again the quantum mechanical formalism is incapable of expressing this dynamical information.

The discussion of the PCM${}^{C}$ in the previous section showed that the configuration space wave function had to be constructed on the configuration sequence tree and could not be constructed from any single process action (sequence tree path). The problem is that the global $\mathcal{H}(\mathcal{M})$ interpretation

$$\hat\Psi_{p}(\mathbf{z}_{1},\ldots,\mathbf{z}_{n})=\sum_{(n^{1}_{i},\ldots,n^{n}_{i})\in \mathcal{I}'}\Gamma_{n^{1}_{i}}\cdots\Gamma_{n^{n}_{i}}T_{\mathbf{m}_{n^{1}_{i}}}g(\mathbf{z}_{1})\cdots T_{\mathbf{m}_{n^{n}_{i}}}g(\mathbf{z}_{n})$$

\noindent suffers from the defect that 

$$T_{\mathbf{m}_{n^{1}_{i}}}g(\mathbf{z}_{1})\cdots T_{\mathbf{m}_{n^{n}_{i}}}g(\mathbf{z}_{n})=0$$

\noindent whenever $(n^{1}_{i},\ldots,n^{n}_{i})\notin \mathcal{I}'$. Therefore, for any configuration of informons that can be generated by the process $\mathbb{P}$ during some action, but not this current action, $\hat\Psi_{p}(\mathbf{z}_{1},\ldots,\mathbf{z}_{n})=0$, which would mean that the configuration should not occur, when in fact it does, and this result will clearly disagree with the quantum mechanical configuration space wave function. The global $\mathcal{H}(\mathcal{M})$ interpretations based on maximal causal tapestries will yield the correct configuration space wave function in the asymptotic limit, but then it becomes clear that this function can no longer provide information about single process actions but only a kind of summarization of all possible process actions. 
Again, information provided by the process model is lost when moving to the quantum mechanical perspective.
 
The Hilbert space algebra itself lacks the nuanced richness of the process algebra. To illustrate this let us briefly consider the case of the two slit experiment and superpositions. 
For simplicity consider a superposition of two distinct states of the same process, $\mathbb{P}=w_{1}\mathbb{P}_{2}\oplus w_{2}\mathbb{P}_{2}$. From the previous section, the exclusive sum is used because the characters of $\mathbb{P}_{1}$ and $\mathbb{P}_{2}$ agree but their values differ. This formulation means that  during a given round only \emph{one} of the $\mathbb{P}_{i}$ is active, generating a single informon. The choice of subprocess may change from round to round, but any informon is generated \emph{only} by a single subprocess. Moroever, because this is an exclusive sum, a single informon will correspond to a single subprocess, either $\mathbb{P}_{1}$ or $\mathbb{P}_{2}$ but never both. Thus every informon corresponds to a single ontological state. There is never any superposition of informons. Thus the causal tapestry $\mathcal{I}$ generated by an action of $\mathbb{P}$ can be properly partitioned into disjoint sets $\mathcal{I}_{1}$ and $\mathcal{I}_{2}$ where $\mathcal{I}_{1}$ consists of the informons generated by $\mathbb{P}_{1}$ and likewise for $\mathcal{I}_{2}$. The global $\mathcal{H}(\mathcal{M})$ interpretation takes the form

$$\Phi_{\mathbb{P}}(\mathbf{z})=\sum_{n\in \mathcal{I}}\Gamma'_{n}T_{\mathbf{m}_{n}}g(\mathbf{z})=\sum_{n_{1}\in\mathcal{I}_{1}}w_{1}\Gamma_{n_{1}}T_{\mathbf{m}_{n_{1}}}g(\mathbf{z})+\sum_{n_{2}\in\mathcal{I}_{2}}w_{2}\Gamma_{n_{2}}T_{\mathbf{m}_{n_{2}}}g(\mathbf{z})=$$

$$w_{1}\Phi_{\mathbb{P}_{1}}(\mathbf{z})+w_{2}\,\Phi_{\mathbb{P}_{2}}(\mathbf{z})$$

\noindent The informons of the subprocesses interleave throughout the causal tapestry and the interpolation allows for the respective global $\mathcal{H}(\mathcal{M})$ interpretations to be formed independently. In the asymptotic limit it is clear that the appropriate quantum mechanical superposition will arise. Note though that the quantum mechanical wave function

$$\Psi(\mathbf{z})=w_{1}\Psi_{1}(\mathbf{z})+w_{2}\,\Psi_{2}(\mathbf{z})$$

\noindent does not distinguish elements supporting the processes separately and this leads to conceptual confusion since it suggests that individual spacetime elements bear properties of both processes simultaneously. The process model eliminates this conceptual confusion because the process algebra is rich enough to keep these two processes ontologically separate. It also shows how important dynamical information is lost is moving over the quantum mechanical perspective.

The two slit case illustrates the subtle distinction between the exclusive and free sums. In the standard quantum mechanical argument, the wave function $\Psi(\mathbf{z})$ of a particle in a two slit experiment is usually divided into two wave functions $\Psi_{L}(\mathbf{z}),\Psi_{R}(\mathbf{z})$ representing the passage of the particle through one of the two slits, say Left and Right. The combined wave function is given as $$\Psi(\mathbf{z})=\frac{1}{\sqrt{2}}(\Psi_{L}(\mathbf{z})+\Psi_{R}(\mathbf{z}))$$

The essential aspect of this wave function is that contributions from the two slits sum to give the correct probability at intermediate locations. From the process perspective, the two slits serve as boundary conditions with which the original generating process $\mathbb{P}$ must interact. These boundary conditions do not result in any change in the character of process nor in its state. Thus when we subdivide $\mathbb{P}$ into subprocesses $\mathbb{P}_{L}$ and $\mathbb{P}_{R}$ we are not creating fundamentally new processes, merely artificially constraining the original process. As a consequence it is appropriate to use the free sum and to write

$$\mathbb{P}=\frac{1}{\sqrt{2}}(\mathbb{P}_{L}\hat\oplus\,\mathbb{P}_{R})$$

 If we examine the system at the informon level then, using the notation of Section IV, the global $\mathcal{H}(\mathcal{M})$ interpretation will have the form

$$\Phi^{m+1}(\mathbf{z})=\sum_{n'\in \mathcal{I}_{m+1}}\Phi_{n'}(\mathbf{z})=\sum_{n'\in \mathcal{I}_{m+1}}\Gamma_{n'}A^{3}T_{\mathbf{m}_{n'}}sinc_{t_{P},l_{P}}(\mathbf{z})=$$

$$\sum_{n'\in \mathcal{I}_{m+1}} \sum_{n\in L_{n'}}\mathcal{T}_{n'n}A^{3}T_{\mathbf{m}_{n'}}sinc_{t_{P},l_{P}}(\mathbf{z})=$$

$$\sum_{n'\in \mathcal{I}_{m+1}}\{ \sum_{n\in L^{L}_{n'}}\mathcal{T}_{n'n}A^{3}T_{\mathbf{m}_{n'}}sinc_{t_{P},l_{P}}(\mathbf{z})\}+$$
$$\sum_{n'\in \mathcal{I}_{m+1}}\{ \sum_{n\in L^{R}_{n'}}\mathcal{T}_{n'n}A^{3}T_{\mathbf{m}_{n'}}sinc_{t_{P},l_{P}}(\mathbf{z})\}=$$

$$\sum_{n'\in \mathcal{I}_{m+1}}{}_{L}\Phi^{m+1}_{n'}(\mathbf{z})+{}_{R}\Phi^{m+1}_{n'}(\mathbf{z})=\Phi^{m+1}_{L}(\mathbf{z})+\Phi^{m+1}_{R}(\mathbf{z})$$

\noindent where $L^{L}_{n'}$ consists of all informons in the $L$ partition of  $\mathcal{I}_{m}$ and ${}_{L}\Phi^{m+1}_{n'}(\mathbf{z})$ is the local $\mathcal{H}(\mathcal{M})$ interpretation at the informon $n'$ determined using information from the $L$ partition. The $R$ case is analogous. $\Phi^{m+1}_{L}$ and $\Phi^{m+1}_{R}$ are the global $\mathcal{H}(\mathcal{M})$ interpretations corresponding to processes $\mathbb{P}_{L}$ and $\mathbb{P}_{R}$ respectively.

The use of the free sum allows single informons to incorporate information from both subprocesses into their generation. This is appropriate since the subprocesses represent artificially distinguished aspects of the same ontological state and so cooperate to generate that same ontological state. Thus their information merely supports the creation of informons corresponding to the \emph{same} ontological state. This is unlike the case of superposition discussed above in which the subprocesses represent \emph{different} ontological states.

We now to turn to the case for products of processes. Consider first the case of two independent particles. In the simplest case the two processes $\mathbb{P}_{1},\mathbb{P}_{2}$ may conjoin as exclusive products

$$\mathbb{P}=\mathbb{P}_{1}\otimes \mathbb{P}_{2}$$.

In this case informons are generated simultaneously as befits the picture of two simultaneously manifesting particles and, moreover, no informon ever incorporates information from both particles so each informon represents an ontologically distinguishable state. The exclusive product was previously described as being fermionic-like due to this non-superposition aspect. This is a reasonable requirement when attempting to model classical processes. In the actual case of fermions the issue is ensuring that the two particles never possess the same state, which would suggest that one could not create couplings of a single fermionic process with itself, i.e. couplings such as

$$\overbrace{\mathbb{P}\hat\otimes\cdots \hat\otimes\mathbb{P}}^{N}\text{  or }\overbrace{\mathbb{P}\otimes\cdots \otimes\mathbb{P}}^{N}$$
 
 \noindent do not exist. This would translate as

$$\overbrace{\mathbb{P}\hat\otimes\cdots \hat\otimes\mathbb{P}}^{N}=\overbrace{\mathbb{P}\otimes\cdots \otimes\mathbb{P}}^{N}=\mathbb{O}$$

In the case that one has distinct fermionic processes it is still reasonable to expect that they should couple via the exclusive sum.

In the case of bosons there is no exclusion principle and so it would seem reasonable to allow either product but it seems most appropriate to use free products for couplings of identical bosonic processes and to use exclusive products for couplings of distinguishable bosonic products.
 This a matter which requires further study.

Entanglement provides an example of an interactive coupling. Consider the case of two scalar particles $A,B$, each of which can be in either of two states $0,1$. In NRQM the wave function of an entanglement of these might take the form

$$\Psi(\mathbf{x}_{A},\mathbf{x}_{B})=\frac{1}{\sqrt{2}}(\Psi^{A}_{0}(\mathbf{x}_{A})\Psi^{B}_{0}(\mathbf{x}_{B})+\Psi^{A}_{1}(\mathbf{x}_{A})\Psi^{B}_{1}(\mathbf{x}_{B}))$$

There are four subprocesses acting here, $\mathbb{P}^{A}_{0},\mathbb{P}^{A}_{1},\mathbb{P}^{B}_{0},\mathbb{P}^{B}_{1}$, one corresponding to each particle-state combination.
These processes are coupled in such a manner that an informon of $\mathbb{P}^{A}_{0}$ appears exclusively with an informon of process $\mathbb{P}^{B}_{0}$ and an informon of $\mathbb{P}^{A}_{1}$ with one of $\mathbb{P}^{B}_{1}$. This means that in the configuration sequence tree, edges will also divide into groups corresponding to these pairings. Thus the global $\mathcal{H}(\mathcal{M})$ interpretation will take the general form 

$$\Phi(\mathbf{z})=\sum_{(n_{1},n_{2})\in \mathcal{I}^{A}_{0}\times \mathcal{I}^{B}_{0}}\negthickspace\Gamma_{n_{1}}\Gamma_{n_{2}}T_{\mathbf{m}_{n_{1}}}sinc_{t_{P}l_{P}}(\mathbf{z})T_{\mathbf{m}_{n_{2}}}sinc_{t_{P}l_{P}}(\mathbf{z}) +$$
$$
\sum_{(n_{3},n_{4})\in \mathcal{I}^{A}_{1}\times \mathcal{I}^{B}_{1}}\negthickspace\Gamma_{n_{3}}\Gamma_{n_{4}}T_{\mathbf{m}_{n_{3}}}sinc_{t_{P}l_{P}}(\mathbf{z})T_{\mathbf{m}_{n_{4}}}sinc_{t_{P}l_{P}}(\mathbf{z})$$

\noindent which is merely one of the global $\mathcal{H}(\mathcal{M})$ interpretations in the configuration process covering map for the coupled process having the form

$$\mathbb{P}=\mathbb{P}^{A}\boxtimes\mathbb{P}^{B}=\frac{1}{\sqrt{2}}((\mathbb{P}^{A}_{0}\otimes\mathbb{P}^{B}_{0})\oplus(\mathbb{P}^{A}_{1}\otimes\mathbb{P}^{B}_{1})$$

The Schr\"odinger cat thought experiment provides another more complex example of an interactive product. In this problem a living cat is placed in a sealed shuttered room with a cannister of cyanide which is released after a radioactive decay takes places. The standard treatment in NRQM is to construct a combined wave function of the form

\begin{displaymath}
\Psi_{C,D} = \frac{1}{\sqrt{2}}[\Psi(D_{n}) \Psi(C_{a}) + \Psi(D_{r}) \Psi(C_{d})]
\end{displaymath}

\noindent where $C,D$ refer to the cat and detector respectively, $D_{n}$ means nonreleased cannister, $C_{a}$ means alive cat, $D_{r}$ means released cannister and $C_{d}$ means dead cat.

If we convert this standard NRQM formulation into process terms, it would take the form
\begin{displaymath}
\mathbb{P}_{C,D} = \frac{1}{\sqrt{2}}[(\mathbb{P}(D_{n})\otimes \mathbb{P}(C_{a})) \oplus (\mathbb{P}(D_{r}) \otimes \mathbb{P}(C_{d}))]
\end{displaymath}

The use of the independent sum implies that on any given step it is possible for either subprocess product to act. In such a case the cat would appear to oscillate randomly between a state of being alive and a state of being dead, or as some would have it, in a weird combination of both.

The problem, however, is that it is impossible for the cat to ever effect a transition from the dead state to the alive state. It can remain indefinitely (more or less if the observer doesn't wait too long) in either the alive or dead state, or transition from alive to dead, but never the converse. As a result it is simply impossible  to form a state for the cat such as  $\frac{1}{\sqrt{2}}[\mathbb{P}(C_{a}) \oplus \mathbb{P}(C_{d})]$ on account of these transition rules. Thus the only proper description for the combined state is as an interactive sum $\frac{1}{\sqrt{2}}[\mathbb{P}(C_{a}) \boxplus \mathbb{P}(C_{d})]$. The sequence tree allows repeated play of the alive process but once the dead process gets activated the only allowable actions are of the dead process.

The proper description of the process in the room is therefore

\begin{displaymath}
\mathbb{P}(D)\boxtimes \mathbb{P}(C)=\frac{1}{\sqrt{2}}[(\mathbb{P}(D_{n})\otimes \mathbb{P}(C_{a})) \boxplus (\mathbb{P}(D_{r}) \otimes \mathbb{P}(C_{d}))]
\end{displaymath}    

\noindent Note that if there is ever an action made by the subprocess $\mathbb{P}(D_{r}) \otimes \mathbb{P}(C_{d})$ then the cat has died and there is thus a transition from the coupled process $\mathbb{P}(D)\boxtimes \mathbb{P}(C)$ to the new process $\mathbb{P}(D_{r}) \otimes \mathbb{P}(C_{d})$. Another way to think of this is if an edge in the configuration space sequence tree is an informon pair from $\frac{1}{\sqrt{2}}[(\mathbb{P}(D_{n})\otimes \mathbb{P}(C_{a})) \boxplus (\mathbb{P}(D_{r}) \otimes \mathbb{P}(C_{d}))]$ then the edge on the next round will be from the same process only if the current informon is restricted to $\frac{1}{\sqrt{2}}(\mathbb{P}(D_{n})\otimes \mathbb{P}(C_{a})) $ but if the current informon is from $\frac{1}{\sqrt{2}}(\mathbb{P}(D_{r}) \otimes \mathbb{P}(C_{d}))$ then the next informon must also be from $\frac{1}{\sqrt{2}}(\mathbb{P}(D_{r}) \otimes \mathbb{P}(C_{d}))$. Note that this does not have a simple algebraic expression in the process algebra and must be described via the configuration sequence tree, a situation alluded to in a previous section.

Many of the paradoxes in NRQM
arise from the dichotomy posed by the distinct ontological properties of particle and of wave.  Recall Bohr's comments on this question \cite{Bohr}

\begin{quote}
... how flawed the simple wave-particle description is. Once light [or a material particle] is in an interferometer, we simply cannot think of it as either a wave or a particle. Nor will melding the two descriptions onto some strange hybrid work. All these attempts are inadequate. What is called for is not a composite picture of light, stitched together out of bits taken from various classical theories. Rather we are being asked for a new concept, a new point of view that will be fundamentally different from those developed from the world of classical physics.
\end{quote} 

The Copenhagen interpretation of NRQM developed by Bohr and his followers proposed a decidedly anti-realist conception of reality. 
Underlying this interpretation is the assumption that a quantum system must be \emph{either} particle-like or wave-like and \emph{not} anything else. But this is a problem with the manner in which such behaviour is represented mathematically.  Both viewpoints represent idealizations and extremes - particles having no spatio-temporal extension while waves have complete spatio-temporal extension.

Contrary to Bohr's unduly pessimistic view, there is in fact a middle ground given by process theory. Representing an entire function $f$ using an interpolation expansion  of the form 

\begin{displaymath}
f(x)=\sum_{x_{i}}f(x_{i})T_{x_{i}}sinc_{\omega}(x)
\end{displaymath}

\noindent neatly incorporates both discrete features, arising from the discrete nature of the sampling set $\{x_{i}\}$, and continuous features, arising from the coupling to the sinc wavelets and the global summation.

The process model resolves the wave-particle duality problem by incorporating one additional aspect - namely it imposes a dynamic on the creation of these interpolation samples such that they are generated sequentially and not simultaneously. As a consequence, at each step in the generation process there is a single localized expression of the quantum system - a discrete, particle-like entity, but due to the extremely small scale at which these entities, these actual occasions manifest, they are unobservable to the emergent entities that form our observable reality (although they are observable to the processes that generate them) and as a result it is only the global $\mathcal{H}\mathcal{M})$-interpretation that is observable, and that interpretation is an interpolation of a continuous, wave-like entity. The process model eliminates the false dichotomy between particle and wave. In the process model, each informon possesses both particle and wave aspects. The particle aspects are represented by the embedding into the causal manifold, which interprets the informon as being associated with a specific space-time location. The wave aspects are represented by the local $\mathcal{H}(\mathcal{M})$-interpretation, which interprets each informon as being a fuzzy wave like entity whose intensity is highly concentrated around the embedding point. The local $\mathcal{H}(\mathcal{M})$-interpretation serves as a frame element contributing to a global interpolation of a wave function over the space-time hypersurface into which the informons embed. Both the embedding and the $\mathcal{H}(\mathcal{M})$-interpretation are emergent aspects of the quantum system.  
  
The subtleties provided by the process algebra allow informons to correspond to distinct, unique ontological states even while the global $\mathcal{H}(\mathcal{M})$ interpretation weaves these disparate states into a coherent whole. This is a property of process and not of the ultimate reality which process generates. Thus one maintains a realist ontology even though the probabilistic structure is that given by NRQM. The emergent probability generated by the process model is inherently non-Kolmogorov. This is illustrated by the two slit case, where informons that can receive information from both subprocesses $\mathbb{P}_{L}$ and $\mathbb{P}_{R}$ will contain tokens of the form $\sum_{i}w^{L}_{i}+ \sum_{j}w^{R}_{j}$ so that in calculating the strength one obtains values of the form 

$$\sum_{i}(w^{L}_{i})^{*}w^{L}_{i} +\sum_{j}(w^{R}_{j})^{*}w^{R}_{j}+\sum_{i}\sum_{j}(w^{L}_{i})^{*}w^{R}_{j} +(w^{R}_{j})^{*}w^{L}_{i}$$

\noindent This is clearly non-Kolmogorov \cite{Khrennikov,Sulis2}.

Note that information flows only in a causally local manner from prior informons to nascent informon and information never flows between informons as they are generated by a process. In the construction of a causal tapestry, information therefore flows only from the prior tapestry and never within the current tapestry as it is being generated. Thus there is, at least in principle, no conflict with relativity, no action at a distance. Of course it will be necessary to construct a proper process model of relativistic quantum mechanics and quantum field theory to be sure that this holds up. 

The process model thus presents an ontology in which the generation of single informons is governed by causally local information. At the process level, however, there is a limited form of nonlocality which arises for two reasons: 1) the fact that processes are generators of space-time but do not possess space-time structure in themselves and 2) the action of process in which successive informons need not be spatio-temporally local to one another, though this does not involve the transfer of information. Non-locality as observed at the measurement level is an emergent non-locality which arises because of the nature of process interactions, especially the interactive coupling, which destroys the statistical independence of the conjoined processes, and because of the measurement situation, which itself is a process interaction.
This weak form of non-locality is termed \emph{quasi-non-locality.}

The process model also possesses a limited form of non-contextuality in that informons are assigned a definite, though limited, set of properties which are inherited from their generating process. It is important to bear in mind that these properties are not directly observable and can only be revealed through an interaction between the generating system and an appropriate measurement process. Moreover, it is not possible for an informon to be assigned a complete set of properties because the generating processes themselves are not capable of supporting complete sets of properties. This is a consequence of the fact that process concatenation is non-commutative as noted previously.
This weak form of non-contextuality is termed \emph{quasi-non-contextuality.}

In the process approach, NRQM appears as an idealization under the asymptotic limit $N,r\rightarrow \infty$ and $t_{P},l_{P}\rightarrow 0$. The latter is the usual limit under which classicality is held to arise. Within the process approach it is postulated that the origin of classicality  arises from the process algebra itself, which allows for the appearance of superselection rules in the form of a null subalgebra consisting of processes that combine as 

$$\mathbb{P}_{1}\oplus\mathbb{P}_{2}=\mathbb{O}\text{ or }\mathbb{P}_{1}\otimes\mathbb{P}_{2}=\mathbb{O}$$
In NRQM it is simply assumed that the linearity of the Schr\"odinger equation implies that any two solutions $\Psi_{1}$ and $\Psi_{2}$ may be summed to give an ontologically realizable state. However the presence of this null sub-algebra means that some combinations might yield the zero process, and so do not generate informons or wave functions at all. Indeed in the Schr\"odinger cat example, $\mathbb{P}(D_{r})\otimes \mathbb{P}(C_{a})=\mathbb{O}$.

How do these super-selection rules arise? Within the process framework, the presence of a sum conjoining two processes $\mathbb{P}_{1},\mathbb{P}_{2}$ implies that they act sequentially. Any sequence of processes is possible, but only one process ever acts during a single round. Staying with the exclusive sum, it is also the case that the two processes never act on the same informon. Generally the exclusive sum is used to represent the situation in which one has a single process type, governed by a single strategy type but possibly where there may be different values available to the properties that may be generated. The individual processes in such a sum are meant to represent different instances of the same process type but with possibly different property values being generated. For example a sum of eigenstates is meant to represent a sum of states for the \emph{same} physical system. The conundrum for classicality is that if we apply the same constraint and assume that in the sum we are representing states of the \emph{same} classical system, then we are faced with asserting that the system exists simultaneously in two distinct classical states, something that simply is never observed.

The way out in the discussion of the Schr\"odinger cat problem was to insist that in the classical setting one must use the interactive sum, rather than the independent sum. But why exactly is this necessary? One approach is to assert that this is a scale phenomenon, not manifesting at quantum mechanical scales but manifesting at classical scales, when $\hslash\rightarrow 0$. In the process model  this limit is necessary to guarantee NRQM as an idealization of the process model (particularly in the case that the wave functions are not bounded in energy and momentum). The same argument cannot be used to obtain classicality at the same time. So another mechanism must be in play.

One thing that distinguishes classical from quantum systems is their size. Another feature is their complexity. A classical system consists of a large number (more often vast number) of components which engage in complex interactions with one another.

Each process taking part in a classical superposition is in fact a complex algebraic tangle of primitive processes, some of which will represent different states of single physical systems. If there are $M$ distinct systems comprising the classical system then we may consider a complex process $\mathbb{P}$ to be an element of $\Pi$ formed from the set $\{\{\mathbb{P}^{1}_{i}\},\{\mathbb{P}^{2}_{j}\},\ldots,\{\mathbb{P}^{M}_{k}\}\}$. If we have a second classical process $\mathbb{Q}$ based upon the same component subsystems (another issue for the Schr\"odinger cat example since a live cat is continually renewing its subsystems, something that a dead cat does not) then it will be an algebraic combination based on the set $\{\{\mathbb{Q}^{1}_{i'}\},\{\mathbb{Q}^{2}_{j'}\},\ldots,\{\mathbb{Q}^{M}_{k'}\}\}$

In the realization of the conjoined process $\mathbb{P}\oplus \mathbb{Q}$ there is no issue until following a round in which an informon of $\mathbb{P}$ is generated, there follows a round in which an informon of $\mathbb{Q}$ is generated. The converse situation can be described analogously. The informon of $\mathbb{P}$ will consist of a collection of informons $\{n^{1},n^{2},\ldots, n^{M}\}$ corresponding to each of the component subprocesses.  

When $\mathbb{Q}$ now acts, the information residing in these new informons, being informons of the same component subsystems as governed by $\mathbb{Q}$, may trigger changes in the subproceses that comprise $\mathbb{Q}$, thus inducing a transition to a new classical process $\mathbb{Q}'$, as in the Schr\"odinger cat example where a transition from $\mathbb{P}(C_{a})\rightarrow \mathbb{P}(C_{d})$ can take place.

It is equally possible that the information, although it does not result in a wholesale change of classical process, may preclude possible moves on the part of $\mathbb{Q}$, so that only a subtree of the subsequent sequence tree may actually be implemented. If such a restriction occurs, then it immediately follows that we are no longer operating within the independent sum $\mathbb{P}\oplus\mathbb{Q}$ but instead have transitioned to the interactive sum $\mathbb{P}\boxplus\mathbb{Q}$. 
At some point in the evolution of these processes there may arise a condition under which it is possible that either $\mathbb{P}$ or $\mathbb{Q}$  be unable to act, and thus rendered inactive, leaving only a single classical process. This may also be the case from the beginning, which would force $\mathbb{P}\oplus\mathbb{Q}=\mathbb{P}\oplus\mathbb{Q}=\mathbb{O}$

The above, though informal, does suggest that the absence of macroscopic superpositions is a expression of the complexity of the conjoining of the individual subprocesses that form a macroscopic or classical object. Indeed, it might be reasonable to \emph{define} a (macroscopic or classical) object to be a complex process for which sums of distinct states are not permitted. In other words, an object would be a collection of complex processes $\{\mathbb{C}_{i}\}$, each generating a distinct state of the object, such that $\mathbb{C}_{i}\oplus\mathbb{C}_{j}=\mathbb{O}$ for all $i,j$.

\section{Conclusion}

The process framework as presented above offers an ontological model for NRQM which is not equivalent to NRQM but rather provides an algebraic and dynamical completion of NRQM. The process framework presents a discrete and finitary model of fundamental reality and NRQM can be viewed as an idealization when the number of fundamental elements and the amount of information transferred between generations can be treated as if infinite. NRQM formally arises though a quotient operation which collapses the richer structure of the process algebra onto the algebraically simpler structure of the Hilbert space. The complexity of the process algebra enables the expression of subtle distinctions between different quantum mechanical situations such as superpositions, multi-slit experiments, entanglement, and perhaps the emergence of classicality. The emergentist approach to spacetime and to physical systems offers a potential resolution to many if not all of the quantum paradoxes which are based on a presumed dichotomy between particle and wave phenomena. In creating the fundamental elements of reality, informons, processes propagate information akin to a discrete wave or diffusion process and the appearance of wave-like or particle-like aspects is a reflection of the choice of measurement process with which the system in question interacts and is not a feature of the informons per se, which are viewed as being quite real, possessing a definite, though incomplete set of properties (quasi-non-conextuality) and whose information is propagated in a causally local manner (quasi-non-locality) even while the processes generating these informons act in an alocal manner \cite{Sulis-thesis}. 

There is no probabilistic structure at the fundamental level. Instead, the probabilistic structures emerges at the measurement level as a consequence of the interactions among processes. This probability structure depends upon the local strength of each process which serves to determine how it couples with other active processes and it is this strength of process which gives the wave-like aspects of informon generation an ontological status.  This in term permits an interpretation of the usual NRQM wave function as an ontological wave manifesting local process strength and the Born rule for interpreting the wave function becomes an emergent rule. This emergent probability structure is non-Kolmogorov in character and conforms to the usual probability structure of NRQM. The connection between the process algebra and NRQM is through an interpolation procedure, which enables calculations to be performed and tested against experiment, an advantage over many other alternative approaches to NRQM.

The basic structure of the process model ensures that information is propagated within any system in a causally local manner and this extends naturally into the relativistic setting. Work needs to be done to see if the process dynamics generally can be extended consistent with relativistic constraints. Should the process approach be successfully extended to relativistic quantum mechanics and quantum field theory, it should avoid many of the problems of those theories associated with divergences. Since the physical entities of the process model are both particle and wave it may be possible to extend the model to field phenomena without the need for second quantization, treating wave phenomena as emergent in the context of large numbers of simpler particles \cite{Mead}.
This in turn would provide a promising new approach to the problem of the unification of quantum mechanics and general relativity.
 \begin{acknowledgments}
Thanks are due to Irina Trofimova for advocating for an emergent, process perspective and to Robert Mann for many fruitful discussions.
\end{acknowledgments}

\end{document}